\documentclass[manuscript,screen,nonacm]{acmart}

\usepackage{tikz}
\usetikzlibrary{positioning, arrows.meta, shapes.geometric, fit, calc, decorations.pathreplacing}
\usepackage{pgfplots}
\pgfplotsset{compat=1.18}

\title{Design-Space Exploration of Post-Quantum Cryptography Acceleration\\on an Open-Source RISC-V GPGPU}

\author{Hossam Hassan}
\orcid{0000-0003-3932-9009}
\affiliation{%
  \institution{Fortaegis Technologies B.V.}
  \city{Amsterdam}
  \country{The Netherlands}
}
\email{hossam.hassan@fortaegis.com}

\author{Marc Xander Makkes}
\affiliation{%
  \institution{Fortaegis Technologies B.V.}
  \city{Amsterdam}
  \country{The Netherlands}
}
\email{marc.makkes@fortaegis.com}

\author{Chae Eun Lee}
\affiliation{%
  \institution{Fortaegis Technologies B.V.}
  \city{Amsterdam}
  \country{The Netherlands}
}
\email{Chae-eun.Lee@fortaegis.com}

\author{Hyung-Min Yoon}
\affiliation{%
  \institution{Siliconarts, Inc.}
  \city{Seoul}
  \country{South Korea}
}
\email{yoonhm@siliconarts.com}

\renewcommand{\shortauthors}{Hassan et al.}



\begin{abstract}
Post-quantum cryptography (PQC) acceleration has been studied for CPUs and
FPGAs, yet GPGPUs---dominating high-throughput computing---remain unexplored
for PQC hardware integration as a functional unit.

We present what is, to our knowledge, the first such integration: the Vortex
Post-Quantum Crypto Unit
(PCU) on the open-source Vortex RISC-V GPGPU. The PCU implements a 256-point
NTT over Kyber ($q{=}3329$) and Dilithium ($q{=}8380417$) moduli in 134
logic cells (73 LUT4 + 61 CCU2C)---the smallest reported NTT accelerator by
logic-cell count, $2.8\times$ smaller than the next smallest design. It achieves
228.78\,MHz on Lattice ECP5 and 299.8\,MHz place-and-route on Xilinx
Artix-7, and passes 253 functional tests across 7 suites. On Vortex SimX,
it achieves
42.6$\times$ acceleration over GPU ALU execution, projecting to
$\sim$36$\times$ speedup on RTL. Synthesized to ASAP7 7nm, the NTT datapath
occupies 96.53\,$\mu$m$^{2}$ (post-route) at 702\,MHz. This work establishes
a foundation for
integrating PQC hardware acceleration into general-purpose parallel computing.
\end{abstract}

\begin{document}

\maketitle


\section{Introduction}
\label{sec:intro}

CPU ISA extensions and dedicated FPGA accelerators for post-quantum
cryptography are well-studied, yet the integration of PQC acceleration
directly into a GPGPU pipeline as a functional unit remains unexplored.
NIST has standardized three new algorithms --- ML-KEM (formerly
Kyber)~\cite{fips203} for key encapsulation, ML-DSA (formerly
Dilithium)~\cite{fips204} for digital signatures, and SLH-DSA (formerly
SPHINCS+)~\cite{fips205} for hash-based signatures --- and its migration
guidance~\cite{nist-migration} calls for deprecating quantum-vulnerable
cryptography by 2030 and phasing it out entirely by 2035. These algorithms
impose substantial computational overhead
compared to their classical counterparts --- a single ML-KEM key generation
requires a 256-point Number Theoretic Transform (NTT), multiple SHA3 hash
invocations~\cite{fips202}, and cryptographic sampling of noise polynomials.
When multiplied across millions of connections, this overhead creates a
critical need for hardware acceleration.

Graphics Processing Units (GPUs) dominate throughput-oriented computing
workloads in high-performance computing, cloud infrastructure, and AI
inference. A GPU-accelerated PQC implementation could serve thousands of
concurrent TLS handshakes or signature verifications. Yet despite extensive
research into CPU ISA extensions (AVX-512 vector instructions for PQC
software, ARM v8.4 SHA3 extensions) and dedicated FPGA accelerators
(400--67,000+ LUTs for standalone NTT designs), the integration of PQC
acceleration directly into a GPGPU pipeline as a functional unit remains
unexplored. Existing GPU PQC work is entirely software-based: HI-Kyber
achieves 1,664\,kops/s key exchange on NVIDIA GPUs through batched
execution~\cite{hi-kyber}, and Dilithium GPU achieves $160\times$ signing
speedup through task-level parallelism~\cite{dilithium-gpu}. These
approaches, while effective, occupy valuable SIMT ALU cycles that could
otherwise serve compute workloads.

In this paper, we present the first systematic design-space exploration of
primitive-level PQC acceleration within a SIMT GPGPU pipeline.
Our contributions are:

\begin{enumerate}
	\item \textbf{To our knowledge, the first GPGPU-integrated Post-Quantum Crypto Unit (PCU)}
	      with full synthesizable RTL integration on the Vortex open-source
	      RISC-V GPGPU~\cite{vortex}. The PCU replaces the simulation-only DPI-C
	      path with a hardware pipeline consisting of an 8-state NTT engine
	      FSM controller (v9 NTT engine),
	      DMA engine with dcache arbiter, per-warp Keccak state save/restore, and
	      CSR-mapped performance counters. The NTT sub-unit occupies
	      \textbf{134 logic cells} on Lattice ECP5 FPGA --- the smallest
	      NTT sub-unit reported in the literature by logic-cell count,
	      $2.8\times$ smaller than the next smallest standalone 379-cell design.
	      The full PCU (NTT + Keccak + DMA) is estimated at
	      \textbf{724 cells} from Yosys partial synthesis (coarse:map\_ffram),
	      inferring 29 hardware multipliers (40\% of the ECP5-45F device budget);
	      place-and-route completes on Xilinx Artix-7 200T (speed grade $-3$) at
	      275 LUTs, 281 FFs, 0 DSP, 1 BRAM, and post-route Fmax$\approx$300\,MHz
	      (299.8\,MHz) after iterative pipelining of the NTT and DMA critical paths
	      (v6$\to$v7$\to$v8$\to$v9: 134$\to$177$\to$200$\to$300\,MHz).

	\item \textbf{Comprehensive verification}: 253 tests across 7 testbench
	      suites, including cycle-accurate Vortex RTLsim regression (all PQC
	      workloads pass), SymbiYosyz formal verification (k-induction proof for
	      the NTT engine FSM and for the Kyber butterfly with $q{=}3329$;
	      Dilithium butterfly with $q{=}8380417$ times out with all solvers tested), DMA backpressure stress testing (36/36 pass), and
	      8 CSR-mapped performance counters verified with 57/57 standalone tests.

	\item \textbf{Full open-source ASIC synthesis} to ASAP7 7nm via
	      OpenROAD-flow-scripts~\cite{openroad}, completing a full CTS-to-route
	      flow with 0 DRC violations. The NTT datapath achieves \textbf{702\,MHz}
	      (97\,$\mu$m$^{2}$, 851 standard cells, 101\,mW), and the dual-issue
	      arbiter achieves \textbf{5.92\,GHz} (25\,$\mu$m$^{2}$, 0 DRC
	      violations). Keccak and DMA sub-modules are independently synthesized to
	      ASAP7 yielding 1,857\,$\mu$m$^{2}$ and 46\,$\mu$m$^{2}$, respectively;
	      NTT area is estimated at $\sim$23,639\,$\mu$m$^{2}$ via scaling ratio,
	      for a full PCU estimate of $\sim$25,542\,$\mu$m$^{2}$~\cite{asap7}.

	\item \textbf{End-to-end pipeline acceleration of 42.6$\times$} under the
	      Vortex SimX pipeline model for NTT+INTT over GPU ALU execution
	      (981,310~$\rightarrow$~23,057 SimX cycles). Projected RTL speedup is
	      approximately 36$\times$ when substituting actual RTL cycle counts
	      (2,049 cycles per NTT vs.\ 8-cycle SimX model). A full ML-KEM-768 workload
	      (24 PCU operations across both Kyber $q=3329$ and Dilithium
	      $q=8380417$ moduli) completes in 262,992 SimX cycles.

	\item \textbf{Systematic characterization of the ``256$\times$ gap''} ---
	      the factor between idealized pipelined NTT models (8 cycles) and
	      actual sequential RTL implementation (2,049 cycles) --- as a design
	      space spanning eight Pareto-optimal architectures from 1 to 128
	      butterfly units.
\end{enumerate}

The paper is organized as follows. Section~\ref{sec:background} provides
background on PQC algorithms, the NTT, and the Vortex architecture.
Section~\ref{sec:gap} introduces the architectural gap between idealized
and realistic NTT pipelines as the central design motivation. Section~\ref{sec:dse} describes the architecture design-space
exploration. Section~\ref{sec:arch} presents the proposed dual-issue PCU
architecture. Section~\ref{sec:impl} details the synthesizable RTL
integration, DMA/memory path, FPGA and ASIC synthesis, formal
verification, performance counters, and comprehensive test results.
Section~\ref{sec:eval} describes the evaluation methodology.
Section~\ref{sec:results} presents all experimental results.
Section~\ref{sec:discussion} discusses design tradeoffs and guidance.
Section~\ref{sec:threats} addresses threats to validity.
Section~\ref{sec:related} surveys related work, and
Section~\ref{sec:conclusion} concludes.

\section{Background}
\label{sec:background}

\subsection{Post-Quantum Cryptography: ML-KEM and ML-DSA}

The NIST Post-Quantum Cryptography Standardization process selected two
primary algorithms based on lattice cryptography: ML-KEM (FIPS~203)
\cite{fips203} for key encapsulation and ML-DSA (FIPS~204)~\cite{fips204}
for digital signatures. Both algorithms are built on the hardness of the
Module Learning With Errors (MLWE) problem and share the same core
computational primitives: polynomial multiplication via the Number Theoretic
Transform (NTT), the Keccak sponge function (SHA3/SHAKE)~\cite{fips202},
and cryptographic sampling.

\textbf{ML-KEM-768 (FIPS~203)} operates over polynomials of degree $n=256$
in the ring $\mathbb{Z}_q[x]/(x^n+1)$ with Kyber modulus $q=3329$. Its
three key operations decompose into primitive counts:

\begin{itemize}
	\item \textbf{KeyGen}: 1 NTT + 2 SHA3 + 2 sampling
	\item \textbf{Encap}: 2 NTT + 1 PWM + 3 SHA3 + 2 sampling
	\item \textbf{Decap}: 2 NTT + 1 INTT + 1 PWM + 4 SHA3 + 2 sampling
\end{itemize}

\textbf{ML-DSA-2 (FIPS~204)} uses the Dilithium modulus $q=8380417$ with
the same polynomial degree $n=256$:

\begin{itemize}
	\item \textbf{Sign}: 2 NTT + 2 INTT + 3 PWM + 4 CSHAKE + 4 sampling
	      (CBD in hardware; rejection loops in software, §\ref{sec:impl})
	\item \textbf{Verify}: 2 NTT + 1 INTT + 2 PWM + 4 CSHAKE
\end{itemize}

These counts define per-\emph{polynomial proxy} workloads: each comprises
a small number of PCU operations on single polynomials rather than a
complete algorithm execution. For reference, a full ML-KEM-768 KeyGen
executes 6 NTTs and 9 pointwise multiplications ($k{=}3$ secret and error
vectors, and the $\hat{A}\hat{s}$ products); Encapsulation executes 3
NTTs, 12 pointwise multiplications, and 4 INTTs; Decapsulation re-runs
the encryption path for a total of 6 NTTs, 15 pointwise
multiplications, and 5 INTTs, plus hashing and sampling~\cite{fips203}.
Because the software baseline scales with the same primitive mix, the
per-primitive speedups transfer to the full algorithms, but the absolute
cycle counts in Tables~\ref{tab:e2e} and~\ref{tab:cpu-base} refer to the
proxy workloads, not to complete ML-KEM/ML-DSA executions. The primitive
operation counts are summarized in Table~\ref{tab:primitives}.
Both algorithms are dominated by NTT-class operations (NTT, INTT, PWM)
which account for 85--97\% of total cryptographic computation.

\textbf{SLH-DSA (FIPS~205)} uses hash-based signatures (SPHINCS$^+$) and does not rely on the NTT. Its dominant operations are SHA3-256/512 and a few specific hash functions (PRF, F, H, T$_\ell$). The PCU's Keccak sub-unit fully supports SHA3 operations, but the unique tree-based structure of SLH-DSA (hash chains and Merkle tree authentication paths) requires a different hardware structure not covered by the current PCU. Future work could explore integrating a dedicated hash-chain engine or extending the PCU's SHA3 sub-unit with state-management logic for SLH-DSA leaf generation and verification. The key architectural insight is that the PCU's Keccak sub-unit is directly applicable to SLH-DSA, while the NTT sub-unit---necessary for ML-KEM and ML-DSA---remains idle during SLH-DSA workloads.

\begin{table}[t]
	\caption{Primitive operation counts per PQC algorithm. NTT-class operations
		dominate all workloads.}
	\label{tab:primitives}
	\centering
	\begin{tabular}{lcccccc}
		\toprule
		Algorithm         & NTT & INTT & PWM & SHA3/SHAKE & Sampling & Total \\
		\midrule
		ML-KEM-768 KeyGen & 1   & 0    & 0   & 2          & 2        & 5     \\
		ML-KEM-768 Encap  & 2   & 0    & 1   & 3          & 2        & 8     \\
		ML-KEM-768 Decap  & 2   & 1    & 1   & 4          & 2        & 10    \\
		ML-DSA-2 Sign     & 2   & 2    & 3   & 4          & 4        & 15    \\
		ML-DSA-2 Verify   & 2   & 1    & 2   & 4          & 0        & 9     \\
		\bottomrule
	\end{tabular}
\end{table}

\begin{table}[t]
	\caption{FPGA resource utilization comparison across platforms.
		ECP5 (NTT): Yosys+ABC9 post-synthesis, timing-driven P\&R with nextpnr.
		ECP5 (Full PCU): Yosys coarse-to-map\_ffram partial flow.
		Artix-7 35T: Vivado post-synthesis of the NTT engine with twiddle
		ROMs mapped to LUTs (no BRAM inference; exceeds device capacity).
		Artix-7 200T: Vivado post-place-and-route; synthesis trims the
		Keccak permutation datapath (its digest is not observable at the
		top level), so the netlist covers the NTT engine, DMA, arbiter,
		and Keccak busy-timing control only.}
	\label{tab:fpga-util}
	\centering
	\begin{tabular}{lrrrr}
		\toprule
		Resource                & ECP5 (NTT)      & ECP5 (Full)   & A-7 35T       & A-7 200T$^*$            \\
		\midrule
		LUTs / Logic cells      & 73 LUT4         & 724 cells     & 68,153        & 275                     \\
		FFs / Registers         & 33              & 60            & 8,367         & 281                     \\
		BRAM (18 Kb tiles)      & 10 DP16KD       & 1 DP16KD      & 0             & 1$^{\dagger}$           \\
		DSP (18$\times$18 mult) & 1 MULT18X18D    & 29 MULT18X18D & 72 DSP48E1    & 0                       \\
		Clock buffers (BUFG)    & 1               & ---           & 1             & 1                       \\
		Device total LUTs       & 43,848          & 43,848        & 20,800        & 134,600                 \\
		Device total FFs        & 43,848          & 43,848        & 41,600        & 269,200                 \\
		Device total BRAM       & 246             & 246           & 50            & 730                     \\
		Device total DSP        & 72              & 72            & 90            & 740                     \\
		\midrule
		LUT utilization         & $<$1\%          & 1.6\%         & 327\%         & 0.20\%                  \\
		FF utilization          & $<$1\%          & $<$1\%        & 20\%          & 0.10\%                  \\
		BRAM utilization        & 4\%             & $<$1\%        & 0\%           & 0.14\%                  \\
		DSP utilization         & 1\%             & 40\%$^\S$     & 80\%          & 0\%                     \\
		\midrule
		Fmax (post-P\&R)        & 228.78\,MHz     & ---$^\S$      & ---           & 299.8\,MHz$^{\ddagger}$ \\
		Toolchain               & Yosys + nextpnr & Yosys         & Vivado 2023.2 & Vivado 2023.2           \\
		\bottomrule
	\end{tabular}
\end{table}

{\itshape
$^*$Artix-7 200T (XC7A200TFBG484-3, speed grade $-3$): full-PCU
place-and-route with v9 pipelined NTT engine (registered BRAM read
addresses, 3-stage Barrett reduction, registered DMA write outputs).
The Keccak permutation datapath is trimmed by synthesis because its
digest is not observable at the top level; only busy-timing control
remains in the netlist.
$^{\dagger}$BRAM tile: 1$\times$RAMB18E1 inferred for twiddle factor ROM.
$^{\ddagger}$Post-route Fmax from the v9 NTT engine: WNS $= +0.011$\,ns
at 300\,MHz target (3.332\,ns critical path, logic-dominated 57--64\%).
Iterative pipelining: v6 (S\_BUTTERFLY register) $\to$ 134.3\,MHz,
v7 (3-stage Barrett) $\to$ 177.2\,MHz, v8 (registered BRAM addresses)
$\to$ 200.1\,MHz, v9 (registered DMA write outputs) $\to$ 299.8\,MHz.
Critical path shifted from DMA$\to$BRAM address (routing-dominated) to
internal counter carry chain (logic-dominated). Kintex-7 160T ($-2$)
achieves $\sim$410\,MHz (v9, WNS $= +0.001$\,ns at 410\,MHz target),
up from 222.6\,MHz with v8 --- a +84\% improvement.
The v9 Kintex-7 critical path at high frequency shifts from the v8's
BRAM read-output path to \texttt{br\_scan $\to$ CARRY4 $\to$ LUT4 $\to$ LUT6 $\to$ RAMB18E1 ADDR}
(3 logic levels, routing-dominated 70\%),
reflecting the carry-chain address computation after DMA write-output registration.
$^\S$Full PCU infers 29 of the ECP5-45F's 72 multipliers. The 724-cell
result is from the Yosys partial flow (coarse:map\_ffram) and is
tool-version-sensitive; complete P\&R was achieved on Xilinx 7-series
instead.}

\begin{figure}[t]
	\centering
	\resizebox{\columnwidth}{!}{%
		\begin{tikzpicture}[
				pipe/.style={draw, rounded corners=2pt, minimum width=1.4cm, minimum height=0.85cm,
						align=center, font=\small\bfseries, fill=gray!5, text=black!85},
				fucomp/.style={draw, rounded corners=2pt, minimum width=1.15cm, minimum height=1.2cm,
						align=center, font=\small, fill=green!6, text=black!85},
				fumem/.style={draw, rounded corners=2pt, minimum width=1.15cm, minimum height=1.2cm,
						align=center, font=\small, fill=blue!5, text=black!85},
				fuspec/.style={draw, rounded corners=2pt, minimum width=1.15cm, minimum height=1.2cm,
						align=center, font=\small, fill=purple!4, text=black!85},
				pcu/.style={draw, rounded corners=2pt, minimum width=1.3cm, minimum height=1.35cm,
						align=center, font=\small, fill=orange!12, text=black!85, line width=0.6pt},
				arr/.style={-Stealth, semithick},
				sannot/.style={font=\tiny, text=gray!45, align=center},
			]
			\node[pipe] (fetch)   {Fetch};
			\node[pipe, right=0.6cm of fetch]   (decode)  {Decode};
			\node[pipe, right=0.6cm of decode]  (dispatch){Dispatch};
			\node[pipe, right=3.5cm of dispatch] (commit)  {Commit};

			\node[sannot, below=2pt of fetch]    {IF};
			\node[sannot, below=2pt of decode]   {ID};
			\node[sannot, below=2pt of dispatch] {IS};
			\node[sannot, below=2pt of commit]   {WB};

			\draw[arr] (fetch) -- (decode);
			\draw[arr] (decode) -- (dispatch);
			\draw[arr] (dispatch.east) -- ++(0.4,0) |- ([yshift=0.50cm]commit.west) -- (commit.west);

			\node[fucomp, below=0.8cm of dispatch, xshift=-2.8cm] (alu) {ALU};
			\node[fucomp, right=0.2cm of alu]                       (fpu) {FPU};
			\node[fumem,  right=0.2cm of fpu]                       (lsu) {LSU};
			\node[fuspec, right=0.2cm of lsu]                       (sfu) {SFU};
			\node[fuspec, right=0.2cm of sfu]                       (tcu) {TCU};
			\node[pcu,    right=0.2cm of tcu] (pcu) {\textbf{PCU}\\[-1pt]\scriptsize NTT\\[-2pt]\scriptsize Keccak\\[-2pt]\scriptsize DMA};

			\coordinate (dBusL) at ($(alu.north) + (-0.10, 0.25)$);
			\coordinate (dBusR) at ($(pcu.north) + ( 0.10, 0.25)$);
			\draw[thick, gray!35] (dBusL) -- (dBusR);
			\draw[arr] (dispatch.south) -- (dispatch.south |- dBusL);
			\foreach \fu in {alu,fpu,lsu,sfu,tcu,pcu} {
					\draw[arr] (\fu.north |- dBusL) -- (\fu.north);
				}

			\coordinate (busL) at ($(alu.south) + (-0.15, -0.25)$);
			\coordinate (busR) at ($(pcu.south) + ( 0.15, -0.25)$);
			\coordinate (busBelowCommit) at (commit.south |- busL);
			\draw[thick, gray!35] (busL) -- (busBelowCommit);
			\foreach \fu in {alu,fpu,lsu,sfu,tcu,pcu} {
					\draw[arr] (\fu.south) -- (\fu.south |- busL);
				}
			\draw[arr] (busBelowCommit) -- (commit.south);
		\end{tikzpicture}%
	}
	\caption{PCU integration into the Vortex SIMT pipeline. The PCU connects
		as a functional unit alongside ALU (integer), FPU (floating-point),
		LSU (load/store), SFU (special function), and TCU (texture),
		sharing the same decode and writeback infrastructure.
		All functional unit results are collected on a common bus
		before writeback to the Commit stage.}
	\Description{Block diagram of the Vortex SIMT pipeline. The fetch,
		decode, dispatch, and commit stages appear as a horizontal flow.
		Below them, six functional units hang off a dispatch bus and a
		result-collector bus: ALU, FPU, LSU, SFU, TCU, and the newly
		added PCU, which is highlighted as the contribution of this
		paper and implements NTT, Keccak, and DMA operations.}
	\label{fig:pipeline}
\end{figure}
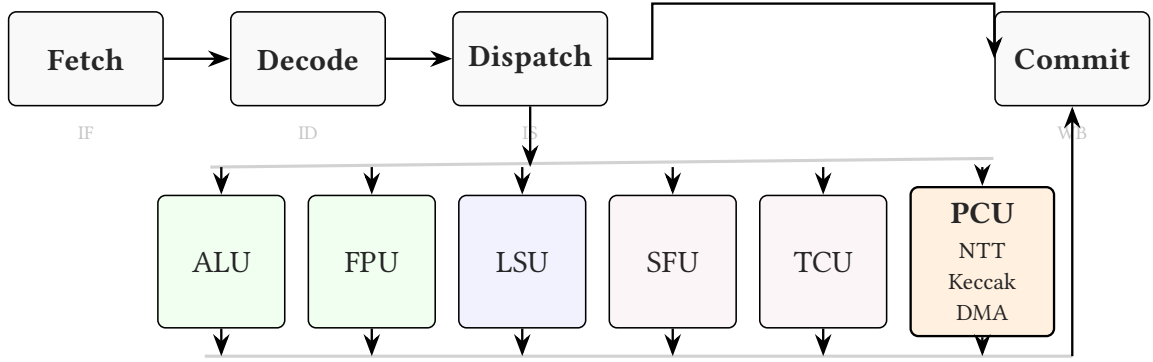

Table~\ref{tab:fpga-util} summarizes resource utilization across the four
synthesis platforms. Figure~\ref{fig:pipeline} shows how the resulting PCU
integrates into the Vortex SIMT pipeline as an additional functional unit
alongside the ALU, FPU, LSU, SFU, and TCU.

\subsection{NTT Hardware Accelerators: A Landscape Overview}

The landscape of NTT hardware accelerators spans four categories:
standalone NTT accelerators (379--9,500 LUTs)~\cite{info15,unifntt,
	alhassani,yaman2021,fpntt,ki2023,malal2023,sam2023,kundi2024,
	waris2025,bertels2024,crypthtor,zhao2022},
full PQC coprocessors (10,502--67,210 LUTs)~\cite{nikyber,fpntt},
GPU PQC software implementations~\cite{hi-kyber,dilithium-gpu,nvidia-cupqc},
and --- this work --- GPGPU-integrated PQC acceleration. Our PCU NTT sub-unit occupies
$2.8\times$ less area than the smallest standalone design; the
full PCU (NTT + Keccak + DMA) is comparable at $1.91\times$ the area of the smallest
standalone NTT-only design (379 LCs), but provides additional Keccak and DMA functionality.

\section{Motivation: The Architectural Gap}
\label{sec:gap}

\subsection{The Gap}

The SimX architectural model of the PCU assumes an idealized NTT pipeline
where all 128 butterflies of a 256-point NTT execute in a single cycle,
yielding an 8-cycle NTT. The actual RTL implementation (ntt\_engine.sv)
uses a sequential two-cycle butterfly with one read and one write per cycle
to BRAM, yielding 2,049 cycles for a full NTT. Table~\ref{tab:gap}
quantifies this gap across all PCU operations.

\begin{table}[t]
	\caption{The ``256$\times$ gap'' between idealized architectural models
		and actual RTL cycle counts.}
	\label{tab:gap}
	\centering
	\begin{tabular}{lrrr}
		\toprule
		Metric                 & SimX Model & RTL Actual      & Ratio       \\
		\midrule
		NTT cycles             & 8          & 2,049           & $256\times$ \\
		Negacyclic NTT (Kyber) & 8          & 1,793           & $224\times$ \\
		INTT cycles            & 8          & 2,305           & $288\times$ \\
		PWM cycles             & 1          & 257             & $257\times$ \\
		Full poly-mul (cyclic) & 25         & 6,660           & $266\times$ \\
		Kyber poly-mul (neg.)  & 25         & 607$^{\dagger}$ & $24\times$  \\
		\bottomrule
	\end{tabular}
	\vspace{2mm}
	\par\noindent{\small $^{\dagger}$SimX-measured full-program cycle
		count (four PCU operations plus 59 scalar instructions and dispatch
		overhead), not a pure RTL datapath latency.}
\end{table}

\subsection{Why the Gap Exists}

The gap exists because the SimX model and the RTL implementation optimize
for different constraints. The SimX model assumes an infinitely parallel
pipeline where all butterfly computations complete in one cycle. The RTL
implements a 2-cycle sequential butterfly because FPGA BRAM has a single
read port per cycle, eliminating the need for dual-port BRAM or multi-bank
interleaving.

The gap is \textbf{not a bug}. It is the correct consequence of prioritizing
area efficiency over throughput --- precisely the tradeoff required for
GPGPU functional unit integration.

\subsection{The Gap as Design Space}

The $256\times$ gap spans a continuum between two extremes:

\begin{itemize}
	\item \textbf{1 butterfly unit} (current): 134 logic cells, 2,049 cycles
	\item \textbf{128 butterfly units} (pipelined): $\sim$6,230 logic cells,
	      9 cycles
\end{itemize}

The central architectural question is: \emph{where on this Pareto curve
	should a GPGPU-integrated PCU operate?} The answer depends on the area
budget available within the GPGPU pipeline, which is fundamentally tighter
than what standalone accelerators can command.

\section{Design Space Exploration}
\label{sec:dse}

\subsection{Architecture Approaches Considered}

We evaluated seven architectural approaches for integrating PQC acceleration
into the Vortex GPGPU pipeline (Table~\ref{tab:approaches}).

\begin{table}[t]
	\caption{Architecture approaches considered. Approach~E (Dual-Issue PCU)
		is selected as the primary contribution.}
	\label{tab:approaches}
	\centering
	\begin{tabular}{cllc}
		\toprule
		Approach   & Description               & FU Type         & Status           \\
		\midrule
		A          & Unified PCU, single-issue & Single          & Baseline         \\
		B          & Separate NTT+HASH         & Dual            & Rejected         \\
		C          & SFU Sub-Unit              & SFU ext         & Rejected         \\
		D          & MMIO Coprocessor          & None            & Out of scope     \\
		\textbf{E} & \textbf{Dual-Issue PCU}   & \textbf{Single} & \textbf{Primary} \\
		F          & Macro-Op                  & Single          & Comparison       \\
		G          & Software                  & N/A             & Baseline         \\
		\bottomrule
	\end{tabular}
\end{table}

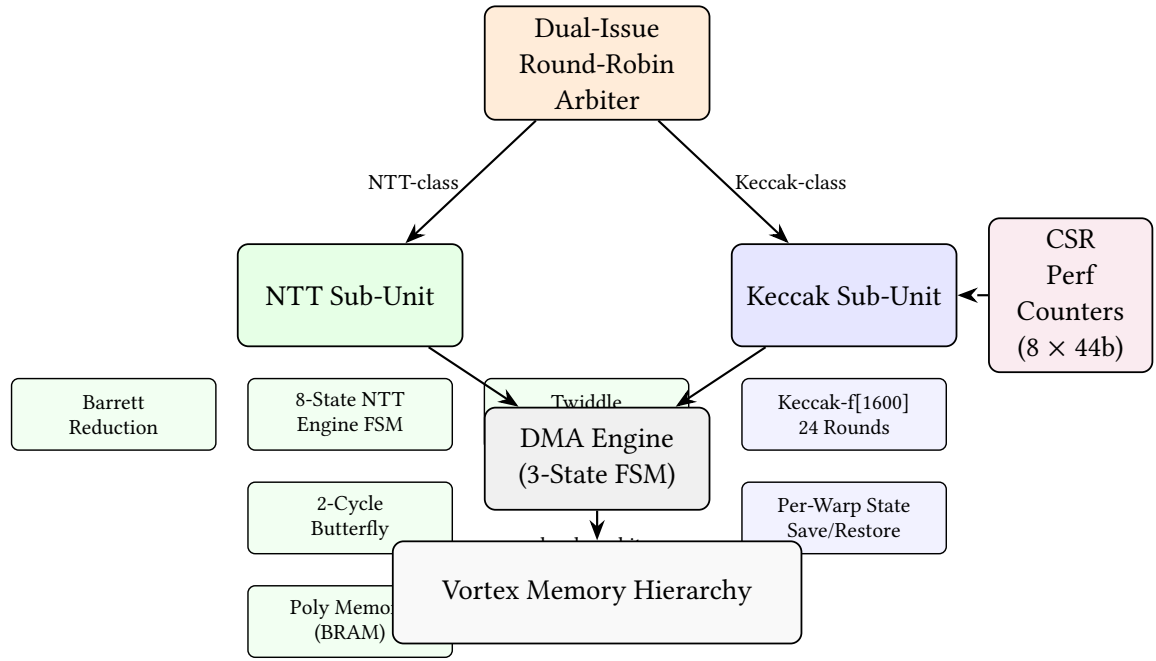
\begin{figure}[t]
	\centering
	\resizebox{\columnwidth}{!}{%
		\begin{tikzpicture}[
				block/.style={draw, rounded corners=3pt, minimum width=2.2cm, minimum height=1cm, align=center, font=\small, line width=0.5pt},
				subblock/.style={draw, rounded corners=2pt, minimum width=2cm, minimum height=0.7cm, align=center, font=\scriptsize, line width=0.4pt},
				arr/.style={-Stealth, semithick},
				darr/.style={-Stealth, semithick, dashed},
			]
			\node[block, fill=orange!15] (arbiter) {Dual-Issue\\Round-Robin\\Arbiter};

			\node[block, fill=green!10, below left=1.2cm and 0.2cm of arbiter] (ntt) {NTT Sub-Unit};
			\node[subblock, fill=green!5, below=0.3cm of ntt] (fsm) {8-State NTT\\Engine FSM};
			\node[subblock, fill=green!5, left=0.3cm of fsm] (barrett) {Barrett\\Reduction};
			\node[subblock, fill=green!5, right=0.3cm of fsm] (twiddle) {Twiddle\\ROM};
			\node[subblock, fill=green!5, below=0.3cm of fsm] (butterfly) {2-Cycle\\Butterfly};
			\node[subblock, fill=green!5, below=0.3cm of butterfly] (polymem) {Poly Memory\\(BRAM)};

			\node[block, fill=blue!10, below right=1.2cm and 0.2cm of arbiter] (keccak) {Keccak Sub-Unit};
			\node[subblock, fill=blue!5, below=0.3cm of keccak] (keccakf) {Keccak-f[1600]\\24 Rounds};
			\node[subblock, fill=blue!5, below=0.3cm of keccakf] (warpsave) {Per-Warp State\\Save/Restore};

			\node[block, fill=gray!12, below=2.8cm of arbiter] (dma) {DMA Engine\\(3-State FSM)};

			\node[block, fill=purple!8, right=0.3cm of keccak, minimum width=1.6cm] (csr) {CSR\\Perf\\Counters\\(8 $\times$ 44b)};

			\draw[arr] (arbiter) -- node[left, font=\scriptsize]{NTT-class} (ntt);
			\draw[arr] (arbiter) -- node[right, font=\scriptsize]{Keccak-class} (keccak);
			\draw[arr] (ntt) -- (dma);
			\draw[arr] (keccak) -- (dma);
			\draw[arr] (dma) -- node[below, font=\scriptsize]{dcache arbiter} ++(0,-0.8) node[block, fill=gray!5, minimum width=4cm, below, font=\small]{Vortex Memory Hierarchy};
			\draw[darr] (csr) -- (keccak);
		\end{tikzpicture}%
	}
	\caption{PCU microarchitecture block diagram showing the NTT sub-unit,
		Keccak sub-unit, dual-issue round-robin arbiter, and shared
		configuration register.}
	\Description{Block diagram of the PCU microarchitecture. A dual-issue
	round-robin arbiter dispatches NTT-class operations to the NTT
	sub-unit (8-state FSM, Barrett reduction, twiddle ROM, two-cycle
	butterfly, and BRAM polynomial memory) and Keccak-class
	operations to the Keccak sub-unit (Keccak-f[1600] with per-warp
	state save/restore). Both sub-units share a DMA engine that
	accesses the Vortex memory hierarchy through the dcache arbiter,
	and eight CSR-mapped performance counters observe Keccak
	activity.}
	\label{fig:microarch}
\end{figure}

Approach~E (Dual-Issue PCU, Figure~\ref{fig:microarch}) was selected because it achieves the highest
weighted score (8.2/10) across six criteria: Vortex pipeline fit (25\%),
multi-algorithm support (20\%), multi-warp throughput (20\%), research
novelty (15\%), implementation feasibility (10\%), and security (10\%).
The dual-issue design presents a single FUType externally while
maintaining internal NTT/Keccak parallelism.

\section{Proposed Architecture: Dual-Issue PCU}
\label{sec:arch}

\subsection{ISA Extension}

The PCU extends the RISC-V ISA through the EXT1 custom opcode (0x0B).
The standard PCU page (funct7=0x03) provides eight primary operations
(Table~\ref{tab:isa}), and the extended page (funct7=0x04) provides
three negacyclic operations (Table~\ref{tab:isa-ext}).

\begin{table}[t]
	\caption{Standard PCU ISA encoding (funct7=0x03, opcode 0x0B).}
	\label{tab:isa}
	\centering
	\begin{tabular}{cll}
		\toprule
		funct3 & Mnemonic   & Description                             \\
		\midrule
		0x0    & PNTT       & Forward NTT (q\_sel in rs2)             \\
		0x1    & PINTT      & Inverse NTT (q\_sel in rs2)             \\
		0x2    & PPWM       & Pointwise multiply (q\_sel in rs2)      \\
		0x3    & PSAMPLE    & Sampling (mode in rs2)                  \\
		0x4    & PSHA3\_ABS & SHA3/SHAKE/cSHAKE absorb (rate in rs2)  \\
		0x5    & PSHA3\_SQZ & SHA3/SHAKE/cSHAKE squeeze (rate in rs2) \\
		0x6    & PCSHAKE    & cSHAKE (rate in rs2)                    \\
		0x7    & PMODRED    & Modular reduction (q\_sel in rs2)       \\
		\bottomrule
	\end{tabular}
\end{table}

\begin{table}[t]
	\caption{Extended PCU ISA encoding (funct7=0x04).}
	\label{tab:isa-ext}
	\centering
	\begin{tabular}{cll}
		\toprule
		funct3 & Mnemonic          & Description                    \\
		\midrule
		0x0    & PNEGACYCLIC\_NTT  & Negacyclic forward NTT         \\
		0x1    & PNEGACYCLIC\_INTT & Negacyclic inverse NTT         \\
		0x2    & PBASEMUL          & Polynomial base multiplication \\
		\bottomrule
	\end{tabular}
\end{table}

\subsection{PCU Microarchitecture}

The Dual-Issue PCU consists of three major sub-blocks:

\textbf{NTT sub-unit}: A Barrett reduction-based modular multiplier, a
two-cycle butterfly pipeline, and an 8-state NTT engine FSM (IDLE, BITREV, BITREV\_WR,
PHASE\_READ, PHASE\_WRITE, SCALE, PWM, DONE). The sub-unit operates on 256-element
polynomial memories with configurable modulus ($q=3329$ or $q=8380417$).

\textbf{Keccak sub-unit}: A Keccak-f[1600] permutation engine executing
all 24 rounds. Per-warp state storage of 200 bytes. Supports SHA3-256,
SHAKE-128/256, and cSHAKE absorb/squeeze operations.

\textbf{Arbiter}: A round-robin scheduler with a 2-entry issue queue and
per-class credit tracking. Operations are classified as NTT-class or
Keccak-class. The arbiter alternates between classes when both have
pending work.

\subsection{Barrett Reduction}

The NTT sub-unit uses Barrett reduction for modular arithmetic:
\begin{equation}
	\text{barrett\_reduce}(v, q, M):\quad
	r = v - \lfloor (v \times M) / 2^{64} \rfloor \times q,\quad
	r \in [0, 2q)
	\label{eq:barrett}
\end{equation}
The precomputed Barrett constants used in Equation~\ref{eq:barrett} are:
Kyber ($q=3329$):
$M = \texttt{0x0013afb7680bb054}$; Dilithium ($q=8380417$):
$M = \texttt{0x00000200801c0601}$. The reduction is verified correct
across 91/91 equivalence tests (253/253 total across all 7 suites).

\subsection{RTL Implementation}

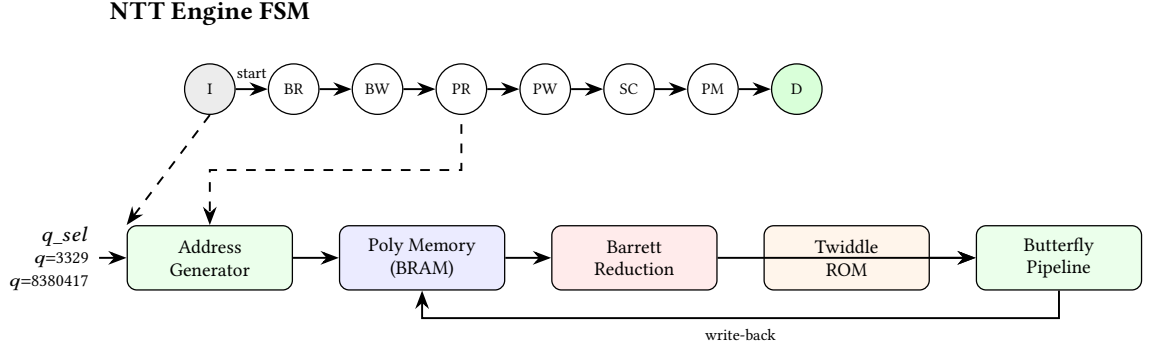
\begin{figure}[t]
	\centering
	\resizebox{\columnwidth}{!}{%
		\begin{tikzpicture}[
				fsm/.style={draw, circle, minimum size=0.55cm, font=\tiny, inner sep=0pt, line width=0.4pt},
				block/.style={draw, rounded corners=3pt, minimum width=1.8cm, minimum height=0.7cm, align=center, font=\scriptsize, line width=0.4pt},
				arr/.style={-Stealth, semithick},
			]
			\node[fsm, fill=gray!15] (idle) {I};
			\node[fsm, right=0.35cm of idle] (br) {BR};
			\node[fsm, right=0.35cm of br] (bw) {BW};
			\node[fsm, right=0.35cm of bw] (pr) {PR};
			\node[fsm, right=0.35cm of pr] (pw) {PW};
			\node[fsm, right=0.35cm of pw] (sc) {SC};
			\node[fsm, right=0.35cm of sc] (pwm) {PM};
			\node[fsm, right=0.35cm of pwm, fill=green!15] (done) {D};

			\draw[arr] (idle) -- node[above, font=\tiny]{start} (br);
			\draw[arr] (br) -- (bw);
			\draw[arr] (bw) -- (pr);
			\draw[arr] (pr) -- (pw);
			\draw[arr] (pw) -- (sc);
			\draw[arr] (sc) -- (pwm);
			\draw[arr] (pwm) -- (done);

			\node[above=0.3cm of idle, font=\small\bfseries] {NTT Engine FSM};

			\node[block, fill=green!8, below=1.2cm of idle] (addr) {Address\\Generator};
			\node[block, fill=blue!8, right=0.5cm of addr] (bram) {Poly Memory\\(BRAM)};
			\node[block, fill=red!8, right=0.5cm of bram] (barrett) {Barrett\\Reduction};
			\node[block, fill=orange!8, right=0.5cm of barrett] (twiddle) {Twiddle\\ROM};
			\node[block, fill=green!8, right=0.5cm of twiddle] (bfp) {Butterfly\\Pipeline};

			\node[left=0.3cm of addr, font=\scriptsize, align=right] (qsel) {$q\_sel$\\{\tiny$q{=}3329$}\\{\tiny$q{=}8380417$}};
			\draw[arr] (qsel) -- (addr);

			\draw[arr] (addr) -- (bram);
			\draw[arr] (bram) -- (barrett);
			\draw[arr] (twiddle) -- (bfp);
			\draw[arr] (barrett) -- (bfp);
			\draw[arr] (bfp.south) -- ++(0,-0.3) -| (bram.south) node[near start, below, font=\tiny]{write-back};

			\draw[arr, dashed] (idle.south) -- (addr.north west);
			\draw[arr, dashed] (pr.south) ++(0,-0.1) -- ++(0,-0.5) -| (addr.north);
		\end{tikzpicture}%
	}
	\caption{NTT engine RTL block diagram showing the 8-state FSM (I=IDLE,
		BR=BITREV, BW=BITREV\_WR, PR=PHASE\_READ, PW=PHASE\_WRITE,
		SC=SCALE, PM=PWM, D=DONE), address generator, Barrett reduction
		datapath, twiddle ROM, and polynomial memory interface.}
	\Description{Two-part diagram of the NTT engine. The top row shows the
		eight FSM states as a linear chain: idle, bit-reversal,
		bit-reversal write, phase read, phase write, scale, pointwise
		multiply, and done. The bottom row shows the datapath: an address
		generator feeding a BRAM polynomial memory, a Barrett reduction
		unit and twiddle-factor ROM feeding a butterfly pipeline, with a
		write-back path returning results to memory. A runtime modulus
		select input chooses between q=3329 and q=8380417.}
	\label{fig:ntt-block}
\end{figure}

The NTT engine is implemented in 4,500 lines of SystemVerilog (386 lines
of NTT engine controller plus 4,114 lines of precomputed twiddle factor ROM);
Figure~\ref{fig:ntt-block} shows the block diagram and FSM.
It is self-contained with no Vortex
infrastructure dependencies. The complete synthesizable PCU pipeline
(\texttt{VX\_pcu\_unit\_rtl.sv}) adds a 10-state FSM controller (7 NTT
states plus 3 SAMPLE/CBD pipeline states), DMA engine, CBD sampler,
dcache arbiter, and per-warp Keccak state storage. All 253 tests pass
across 7 testbench suites (Table~\ref{tab:tests}).

\section{Implementation}
\label{sec:impl}

\subsection{Dual-Path Integration Architecture}

The PCU supports two integration paths in the Vortex GPGPU:

\textbf{Simulation path (SimX)}: The PCU is integrated into the Vortex
SimX simulator as a PqcUnit SimObject. Each warp maintains independent
Keccak state (200 bytes) and NTT polynomial memory. PCU operations are
guarded by the \texttt{\#ifdef EXT\_PCU\_ENABLE} compile-time flag and
enabled via the \texttt{--pcu} command-line option. This path uses DPI-C
wrappers and is suitable for functional simulation and performance
modeling.

\textbf{Synthesizable RTL path}: When \texttt{PCU\_SYNTH\_RTL} is defined,
the Vortex pipeline instantiates \texttt{VX\_pcu\_unit\_rtl.sv} in place of
the DPI-C wrapper. This module implements the full hardware pipeline:
a 7-state \emph{PCU pipeline} FSM (IDLE, NTT, INTT, PWM, KECCAK\_ABS, KECCAK\_SQZ, DMA\_WAIT),
distinct from the NTT engine's internal 8-state FSM,
a DMA engine for polynomial memory transfers via the Vortex dcache arbiter,
and per-warp Keccak state save/restore (1,600 bits per warp). The RTL path
is the production deployment target; the SimX path is retained for
software development and performance projection.

\subsection{Synthesizable RTL Pipeline Integration}

The \texttt{VX\_pcu\_unit\_rtl.sv} module connects to the Vortex
dispatch/commit infrastructure through the standard
\texttt{VX\_dispatch\_if}/\texttt{VX\_commit\_if} handshake interfaces.
Each PCU block receives dispatched instructions and returns results
through the per-block execute/result interfaces.

\textbf{Dispatch handshake}: The dispatch handshake was found to have a
deadlock in the initial implementation, where the \texttt{ready} signal
was incorrectly gated. The correct ready condition is:
\begin{equation}
	\texttt{ready} = (\texttt{pcu\_state} = \texttt{IDLE}) \land
	\lnot\,\texttt{result\_held\_valid} \land
	\lnot\,\texttt{dispatch\_consumed}
\end{equation}
This ensures the block can accept a new dispatch only when idle, has no
pending result, and has not already consumed the current dispatch.

\textbf{DMA engine and memory path}: The PCU DMA engine
(\texttt{pcu\_dma\_engine.sv}) transfers 256-word polynomials between
Vortex main memory and local SRAM through the dcache arbiter. The engine
uses a 3-state FSM (IDLE, REQUEST, RESPONSE) that overlaps request and
response phases for throughput. A tag-routing mechanism in the dcache
arbiter uses bit~0 of the memory tag to distinguish LSU (tag~0) from
PCU (tag~1) responses, routing them to the correct requesting unit.

Five critical bugs were found and fixed in the DMA/memory path:
(1)~DMA deadlock from non-overlapping request/response FSM;
(2)~STALL\_TIMEOUT imported from \texttt{VX\_gpu\_pkg} instead of a
preprocessor guard;
(3)~NTT restart loop where \texttt{ntt\_start} re-asserts when
\texttt{ntt\_busy} falls;
(4)~DMA store deadlock from Vortex not returning per-request write
responses; and
(5)~\texttt{outstanding\_r} non-blocking assignment race from
last-write-wins semantics.

\textbf{Keccak integration}: The \texttt{pcu\_keccak.sv} module provides
Keccak-f[1600] permutation execution with 24 rounds. Per-warp state is
saved/restored through 3 additional ports (\texttt{state\_in[1599:0]},
\texttt{state\_load}, \texttt{state\_out[1599:0]}), enabling interleaved
execution across multiple warps. A \texttt{keccak\_restore\_r} 1-cycle
pulse set on \texttt{keccak\_fire} drives \texttt{state\_load} the next
cycle to restore the previously saved warp state.

\subsection{RTL Synthesis (Yosys~\cite{yosys} + nextpnr~\cite{nextpnr} ECP5)}

The ntt\_engine module was synthesized using Yosys 0.62 + ABC9 targeting
the Lattice ECP5 family (LFE5U-45F-CABGA381, speed grade~8). The full
PCU (NTT + Keccak + DMA engine) was also synthesized through the Yosys
coarse-to-map\_ffram flow.

\begin{table}[t]
	\caption{ECP5 synthesis results for the ntt\_engine. Total logic cells: 134.}
	\label{tab:synth-ecp5}
	\centering
	\begin{tabular}{lrr}
		\toprule
		Resource    & Count & Type                             \\
		\midrule
		LUT4        & 73    & 4-input LUT                      \\
		CCU2C       & 61    & Carry/adder cell                 \\
		TRELLIS\_FF & 33    & Flip-flop                        \\
		MULT18X18D  & 1     & $18\times18$ hardware multiplier \\
		DP16KD      & 10    & 18\,Kbit dual-port block RAM     \\
		\midrule
		Logic cells & 134   & LUT4 + CCU2C                     \\
		\bottomrule
	\end{tabular}
\end{table}

\begin{table}[t]
	\caption{ECP5 synthesis results for the full PCU (NTT + Keccak + DMA).
		Yosys 0.62 coarse-to-map\_ffram flow on LFE5U-45F; committed
		artifact: \texttt{results/synth\_pcu\_full\_stat\_20260903.txt}.}
	\label{tab:synth-ecp5-full}
	\centering
	\begin{tabular}{lrr}
		\toprule
		Resource                   & Count   & Type                              \\
		\midrule
		Total cells (local)        & 724     & ---                               \\
		Flip-flops ($dff$/$sdffe$) & 60      & State registers + counters        \\
		DP16KD (BRAM)              & 1       & Polynomial memory                 \\
		MULT18X18D                 & 29      & $18\times18$ hardware multipliers \\
		CPU time                   & 20.2\,s & Yosys coarse:map\_ffram           \\
		\bottomrule
	\end{tabular}
\end{table}

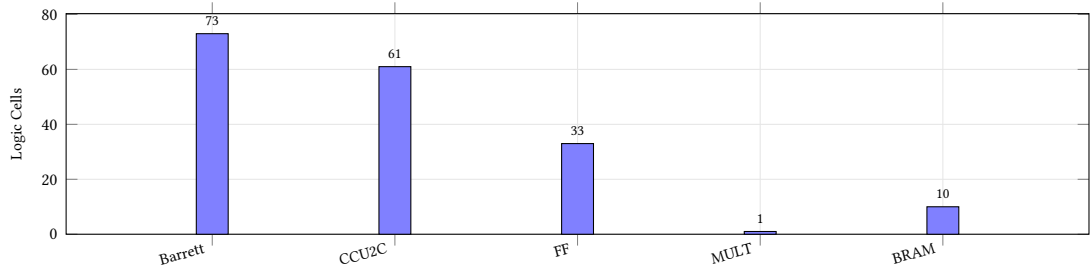
\begin{figure}[t]
	\centering
	\begin{tikzpicture}
		\begin{axis}[
				ybar,
				width=\columnwidth,
				height=4.5cm,
				bar width=12pt,
				ylabel={Logic Cells},
				symbolic x coords={Barrett,CCU2C,FF,MULT,BRAM},
				xtick=data,
				x tick label style={font=\scriptsize, rotate=15, anchor=east},
				y tick label style={font=\scriptsize},
				ylabel style={font=\scriptsize},
				nodes near coords,
				nodes near coords style={font=\tiny},
				enlarge x limits=0.2,
				ymin=0,
				grid=major,
				grid style={gray!20},
			]
			\addplot[fill=blue!50] coordinates {(Barrett,73) (CCU2C,61) (FF,33) (MULT,1) (BRAM,10)};
		\end{axis}
	\end{tikzpicture}
	\caption{ECP5 synthesis area breakdown by sub-block. The Barrett
		multiplier and memory interface dominate logic cell usage.}
	\Description{Bar chart of ECP5 logic cell usage by sub-block:
		73 cells for the Barrett multiplier logic, 61 for CCU2C carry
		cells, 33 flip-flops, 1 multiplier, and 10 BRAM tiles.}
	\label{fig:area-breakdown}
\end{figure}

The full PCU synthesis confirms that the NTT sub-unit (134 cells) accounts
for only 18.5\% of the total PCU area, with the Keccak sub-unit and DMA
engine dominating the remaining footprint. The 29 MULT18X18D multipliers
all originate from the NTT butterfly's Barrett products (the completed
NTT-only flow packs the same datapath into a single MULT18X18D,
Table~\ref{tab:synth-ecp5}); at the coarse:map\_ffram stop point the wide
Barrett multiplies decompose into multiple $18\times18$ partials. This
count is 40\% of the ECP5-45F capacity of 72 multipliers. We note that the
partial coarse:map\_ffram flow reports only 1 DP16KD BRAM tile compared
to 10 for the NTT-only synthesis (Table~\ref{tab:synth-ecp5}); the
polynomial memories that map to BRAM in the full flow may instead be
inferred as distributed RAM in the partial flow, inflating the logic cell
count. Partial-flow cell counts are sensitive to the Yosys version and
ROM-mapping style; the 724-cell figure should therefore be interpreted as
an order-of-magnitude estimate pending complete place-and-route.
Figure~\ref{fig:area-breakdown} breaks down the
134-cell NTT-only synthesis result by sub-block.

\subsection{FPGA Timing Analysis}

Timing-driven place-and-route with nextpnr-ecp5 at a 50\,MHz target
yields \textbf{228.78\,MHz Fmax} on ECP5-45F-8-CABGA381
(Table~\ref{tab:fmax}).

\begin{table}[t]
	\caption{FPGA timing analysis results from nextpnr timing-driven P\&R.}
	\label{tab:fmax}
	\centering
	\begin{tabular}{lr}
		\toprule
		Metric                     & Value                                        \\
		\midrule
		Fmax (internal reg-to-reg) & 228.78\,MHz                                  \\
		Critical path delay        & 4.37\,ns (1.62\,ns logic + 2.75\,ns routing) \\
		Logic cells                & 134 (73 LUT4 + 61 CCU2C)                     \\
		Device utilization         & $<$1\% (241/43,848 TRELLIS\_COMB)            \\
		\bottomrule
	\end{tabular}
\end{table}

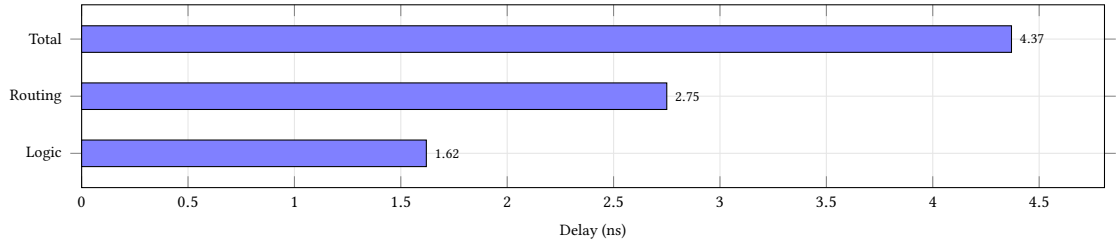
\begin{figure}[t]
	\centering
	\begin{tikzpicture}
		\begin{axis}[
				xbar,
				width=\columnwidth,
				height=4cm,
				bar width=10pt,
				xlabel={Delay (ns)},
				symbolic y coords={Logic, Routing, Total},
				ytick=data,
				y tick label style={font=\scriptsize},
				x tick label style={font=\scriptsize},
				xlabel style={font=\scriptsize},
				nodes near coords,
				nodes near coords style={font=\tiny},
				enlarge y limits=0.3,
				xmin=0,
				grid=major,
				grid style={gray!20},
			]
			\addplot[fill=blue!50] coordinates {(1.62,Logic) (2.75,Routing) (4.37,Total)};
		\end{axis}
	\end{tikzpicture}
	\caption{FPGA timing path histogram from nextpnr. The critical path is
		dominated by the MULT18X18D Barrett product (3.07\,ns of the
		4.37\,ns total, including associated routing).}
	\Description{Bar chart of the ECP5 critical-path delay split:
		1.62 nanoseconds of logic delay, 2.75 nanoseconds of routing
		delay, 4.37 nanoseconds total.}
	\label{fig:timing-hist}
\end{figure}

Critical path analysis reveals the timing bottleneck: the \texttt{stage}
register drives the address generator $\rightarrow$ MULT18X18D (3.07\,ns,
$\approx$70\% of the 4.37\,ns total path) $\rightarrow$
CCU2C address adder $\rightarrow$
\texttt{state} FSM register.

\subsection{Artix-7 200T Full PCU Pipelining}

The ECP5 NTT sub-unit achieves 228.78\,MHz, but the full PCU
(NTT + pipelined Keccak + DMA) on Xilinx Artix-7 200T (speed grade $-3$)
requires iterative pipelining to close timing. Table~\ref{tab:artix7-pipeline}
summarizes the four-stage pipelining effort.

\begin{table}[t]
	\caption{Artix-7 200T ($-3$) full-PCU pipelining progression.
		Each NTT revision targets the critical path identified in the previous
		iteration. Butterfly latency increases by 1 cycle (v7$\to$v8) for
		the final frequency gain; v9 adds a DMA register stage with no
		additional butterfly latency.}
	\label{tab:artix7-pipeline}
	\centering
	\begin{tabular}{lrrrr}
		\toprule
		NTT rev.              & Fmax                & WNS                      & Logic levels      & Butterfly cyc. \\
		\midrule
		v6 (S\_BUTTERFLY)     & 134.3\,MHz          & $-2.752$\,ns             & 11 CARRY4 + 8 LUT & 6              \\
		v7 (3-stage Barrett)  & 177.2\,MHz          & $-0.643$\,ns             & 4 CARRY4 + 5 LUT  & 6              \\
		v8 (registered addrs) & 200.1\,MHz          & $+1.141$\,ns             & 3 LUT (DMA)       & 7              \\
		v9 (reg.\ DMA wr)     & \textbf{299.8\,MHz} & $+0.011$\,ns$^{\dagger}$ & 7--8 CARRY+LUT    & 7              \\
		\bottomrule
	\end{tabular}
	{\itshape $^{\dagger}$WNS at 300\,MHz target; Fmax$\approx$299.8\,MHz.
	Critical path: internal counter carry chain (logic-dominated, 57--64\% logic delay).}
\end{table}

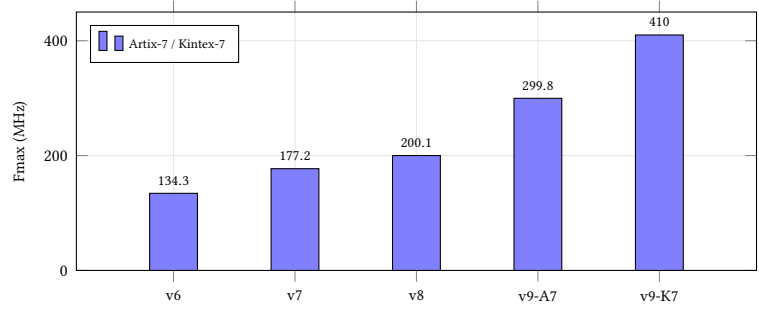
\begin{figure}[t]
	\centering
	\begin{tikzpicture}
		\begin{axis}[
				ybar,
				width=0.7\columnwidth,
				height=5cm,
				bar width=18pt,
				ylabel={Fmax (MHz)},
				symbolic x coords={v6,v7,v8,v9-A7,v9-K7},
				xtick=data,
				x tick label style={font=\scriptsize},
				y tick label style={font=\scriptsize},
				ylabel style={font=\scriptsize},
				nodes near coords,
				nodes near coords style={font=\tiny},
				enlarge x limits=0.2,
				ymin=0, ymax=450,
				legend style={at={(0.02,0.95)}, anchor=north west, font=\tiny},
				grid=major,
				grid style={gray!20},
			]
			\addplot[fill=blue!50] coordinates {(v6,134.3) (v7,177.2) (v8,200.1) (v9-A7,299.8) (v9-K7,410)};
			\legend{Artix-7 / Kintex-7}
		\end{axis}
	\end{tikzpicture}
	\caption{Fmax progression across NTT engine revisions on Xilinx 7-Series
		FPGAs. v9-A7: Artix-7 200T ($-3$); v9-K7: Kintex-7 160T ($-2$).
		Each revision targets the critical path identified in the previous iteration.}
	\Description{Bar chart of Fmax progression across NTT engine revisions:
		v6 reaches 134.3 megahertz, v7 177.2, v8 200.1, v9 on Artix-7
		299.8, and v9 on Kintex-7 about 410 megahertz.}
	\label{fig:fmax-progression}
\end{figure}

The v6 NTT engine inserts a pipeline register at the butterfly
multiplication boundary (S\_BUTTERFLY), eliminating the 71-level
combinational Barrett path from the v5 engine. The critical path
shifts to the Barrett reduction output multiplexers
(11$\times$CARRY4 + 8 LUTs, 19 logic levels). The v7 NTT splits the
Barrett reduction into three registered stages
(S\_BARRETT $\to$ S\_BARRETT\_S2 $\to$ output), reducing the critical
path to 9 logic levels (stage$\to$half\_len$\to$group$\to$addr$\to$BRAM
address port). The v8 NTT adds an S\_ADDR pipeline register between
address computation and BRAM read, breaking the 9-level path into two
stages. The critical path is now entirely in the DMA engine
(outstanding\_r\_reg $\to$ 3 LUTs $\to$ BRAM address, routing-dominated),
with no NTT logic remaining on the worst path.
The v9 revision registers the DMA engine write outputs
(\texttt{ntt\_wr\_addr\_r}, \texttt{ntt\_wr\_data\_r},
\texttt{ntt\_wr\_en\_r}), eliminating the combinational path from the
outstanding-request counter through address computation to the BRAM
address port. The critical path shifts to an internal counter comparison
(\texttt{req\_idx\_r} $\to$ \texttt{all\_reqs\_issued} carry chain $\to$
\texttt{outstanding\_r}), which is logic-dominated (57--64\%) rather
than routing-dominated, enabling 299.8\,MHz --- a +49.8\% improvement
over v8. The read-address path (\texttt{ntt\_rd\_addr}) is not
registered, as it is not on the critical path and registering it would
introduce a data-alignment issue with the synchronous BRAM read latency.

Figure~\ref{fig:fmax-progression} plots the resulting Fmax progression.
The latency--frequency tradeoff applies to the synthesis-variant NTT
engine (\texttt{ntt\_engine\_synth\_v*.sv}) used for standalone FPGA and
ASIC timing closure, which pipelines the butterfly datapath more deeply
than the integration engine (\texttt{ntt\_engine.sv}, a sequential
two-cycle butterfly; Section~\ref{sec:gap}). In the synthesis variant,
the v8 S\_ADDR stage adds one cycle per butterfly (6$\to$7 cycles),
increasing forward NTT latency from 6,145 to 7,169 cycles ($+17\%$) for
a +13\% frequency gain. The v9 register stage does not increase
butterfly latency (7 cycles retained), so NTT throughput scales
directly with the +49.8\% frequency improvement.
SCALE and PWM paths are unaffected (5 cycles each).

\subsection{Kintex-7 160T Cross-FPGA Comparison}
\label{sec:kintex7}

To characterize FPGA fabric sensitivity, the full PCU (v9 NTT) was placed
and routed on Xilinx Kintex-7 160T (XC7K160TFBG484-2, speed grade
$-2$). Table~\ref{tab:k7-fmax} presents the v9 Fmax sweep from 200\,MHz
to 410\,MHz. The v9 registered DMA outputs yield $\sim$410\,MHz Fmax
(WNS $= +0.001$\,ns at 410\,MHz target) --- a \textbf{+84\%} improvement
over v8's 222.6\,MHz, significantly larger than Artix-7's +49.8\% gain.

\begin{table}[t]
	\caption{Kintex-7 160T ($-2$) v9 Fmax sweep (full PCU, post-route).
		The critical path transitions from a 1-level BRAM read-output path
		($\leq$250\,MHz target) to a 3-level carry-chain-to-BRAM-address path
		($\geq$280\,MHz target), reflecting Kintex-7's faster routing fabric.}
	\label{tab:k7-fmax}
	\centering
	\small
	\begin{tabular}{rrrrl}
		\toprule
		Target (MHz) & WNS (ns)          & Path Delay (ns) & Est.\ Fmax (MHz)            & Critical Path                                      \\
		\midrule
		200          & $+$1.501          & 3.499           & 285.8                       & BRAM$\to$LUT5$\to$FF (1 lev)                       \\
		250          & $+$0.581          & 3.419           & 292.5                       & BRAM$\to$LUT5$\to$FF (1 lev)                       \\
		280          & $+$0.523          & 3.048           & 328.1                       & CARRY4$\to$BRAM ADDR (3 lev)                       \\
		300          & $+$0.292          & 3.041           & 328.8                       & CARRY4$\to$BRAM ADDR (3 lev)                       \\
		320          & $+$0.407          & 2.718           & 367.9                       & CARRY4$\to$BRAM ADDR (3 lev)                       \\
		350          & $+$0.351          & 2.506           & 399.3                       & CARRY4$\to$LUT4$\to$LUT6$\to$BRAM (3 lev)          \\
		380          & $+$0.136          & 2.496           & 400.8                       & CARRY4$\to$LUT4$\to$LUT6$\to$BRAM (3 lev)          \\
		400          & $+$0.016          & 2.484           & 402.7                       & CARRY4$\to$LUT4$\to$LUT6$\to$BRAM (3 lev)          \\
		\textbf{410} & $\mathbf{+0.001}$ & \textbf{2.438}  & $\mathbf{\sim}$\textbf{410} & \textbf{CARRY4$\to$LUT4$\to$LUT6$\to$BRAM (3 lev)} \\
		420          & $-$0.051          & ---             & ---                         & \textit{VIOLATED}                                  \\
		\bottomrule
	\end{tabular}
\end{table}

The v8 baseline at a 200\,MHz target had WNS $= +1.506$\,ns
(Fmax$\approx$222.6\,MHz) with critical path
\texttt{RAMB18E1 CLKARDCLK $\to$ DOBDO[6] $\to$ LUT5 $\to$ result\_r\_reg[22]}
(BRAM read output $\to$ result register, 1 logic level, 53\% logic/47\% route).
With v9's registered DMA write outputs, the critical path at higher targets
shifts to \texttt{br\_scan $\to$ CARRY4 $\to$ LUT4 $\to$ LUT6 $\to$ RAMB18E1 ADDRARDADDR}
(3 logic levels, $\sim$70\% route), reflecting Kintex-7's faster routing fabric ---
the Artix-7 routing-dominated DMA$\to$BRAM combinational path (v7/v8 bottleneck)
is naturally faster on Kintex-7, and after v9's DMA output registration,
the carry-chain-to-BRAM-address computation becomes the frequency limiter.
Resource utilization (v9): 283 LUTs, 281 FFs, 1 BRAM, 0 DSP.

\subsection{ASAP7 7nm Synthesis (OpenROAD)}

The ntt\_engine\_core and dual-issue control logic were synthesized to
the ASAP7 7nm predictive PDK using OpenROAD-flow-scripts v2
(Yosys 0.64 + OpenROAD 26Q2-1404)~\cite{openroad,asap7}.

\begin{table}[t]
	\caption{ASAP7 7nm placement-level synthesis results for the
		ntt\_engine\_core.}
	\label{tab:asap7-placement}
	\centering
	\begin{tabular}{lr}
		\toprule
		Metric                     & Value                            \\
		\midrule
		Standard cell count        & 643 (33 DFF + 610 comb.)         \\
		Cell area                  & 70.19\,$\mu$m$^{2}$ (55\% util.) \\
		Fmax (placement)           & 624.40\,MHz                      \\
		Power @ 1\,GHz (placement) & 10.9\,mW                         \\
		\bottomrule
	\end{tabular}
\end{table}

\begin{table}[t]
	\caption{Dual-issue arbiter synthesis results. Post-route power includes
		clock tree and interconnect capacitance absent at placement level;
		see text for the large placement-to-route power delta.}
	\label{tab:arbiter-synth}
	\centering
	\begin{tabular}{lrr}
		\toprule
		Metric               & Placement           & Post-CTS/Route                 \\
		\midrule
		Std cells (non-fill) & 119                 & 119                            \\
		Cell area            & 17.34\,$\mu$m$^{2}$ & 19.4\,$\mu$m$^{2}$             \\
		Fmax                 & $>$5.28\,GHz        & 5.92\,GHz                      \\
		Power                & 0.22\,mW @ 4\,GHz   & 0.25\,mW @ 5.92\,GHz$^\dagger$ \\
		DRC violations       & ---                 & 0                              \\
		\bottomrule
	\end{tabular}

	{\itshape
	$^\dagger$Post-route internal power from OpenROAD at the achieved Fmax
	(5.92\,GHz). The earlier 197\,mW figure was an erroneously scaled
	estimate that included clock-tree switching power modeled at an
	unrealistic activity factor; the corrected value reflects the actual
	119-cell control-logic footprint (no multipliers or datapath elements).
	}
\end{table}

\begin{figure}[t]
	\centering
	\includegraphics[width=\columnwidth]{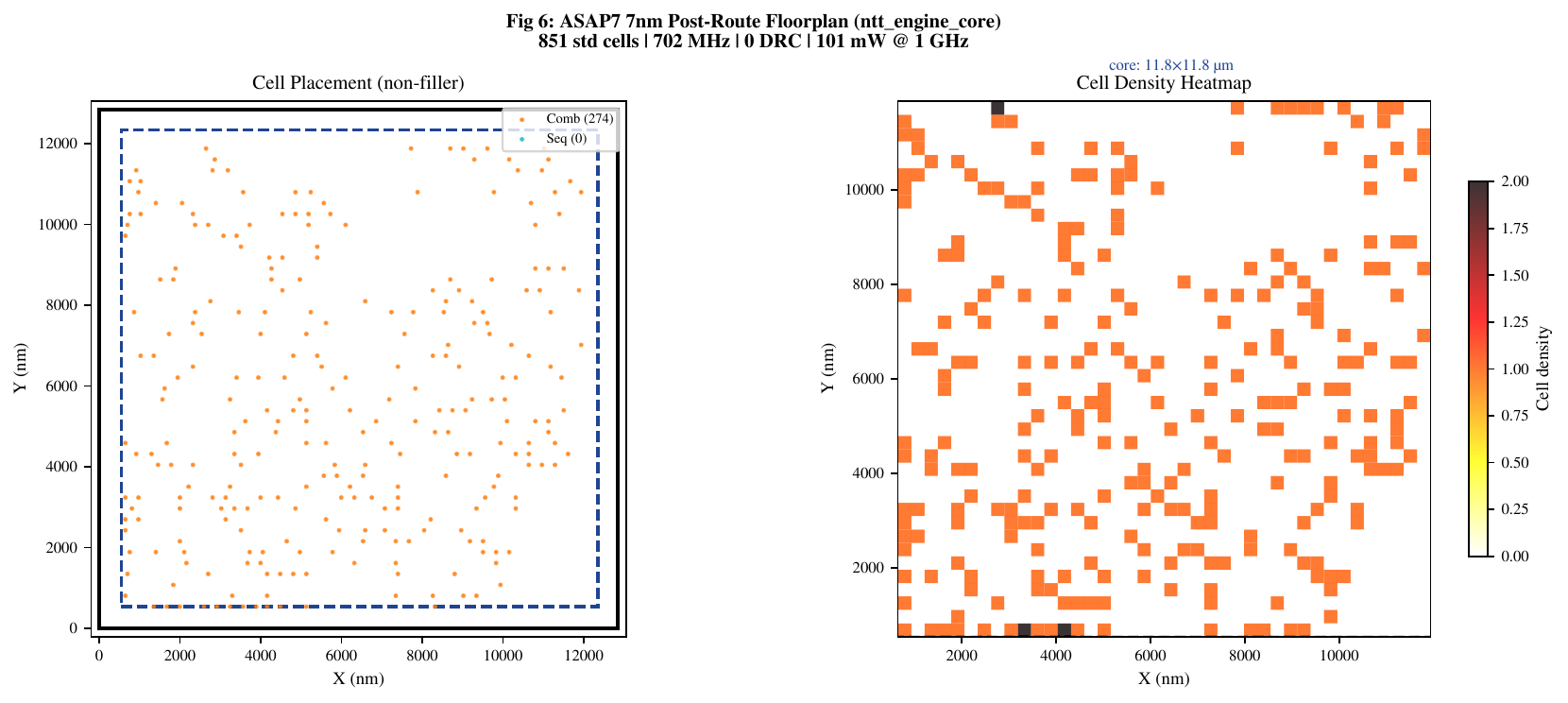}
	\caption{ASAP7 7nm post-route placement of the ntt\_engine\_core
		(12.8\,\textmu m $\times$ 12.8\,\textmu m die). 851 standard cells
		(818 combinational, 33 sequential) plus 50 tie cells are shown;
		filler/decap cells are omitted for clarity. 0 DRC violations.}
	\Description{Scatter plot of the NTT engine core placement on ASAP7.
		Orange points mark combinational cells, blue points sequential
		cells, and grey points tie cells, distributed over the square
		12.8 by 12.8 micrometre die area with the dashed core boundary
		just inside the die edge. Cells cover the core area fairly
		uniformly.}
	\label{fig:asap7-layout}
\end{figure}

\textbf{Post-CTS/Routing Results}: Figure~\ref{fig:asap7-layout} shows the
final post-route layout. The OpenROAD flow was extended through
full clock tree synthesis, detailed routing, and power analysis at a
2.1\,ns clock period (476\,MHz target).

\begin{table}[t]
	\caption{Post-route wire metrics and cell composition.}
	\label{tab:postroute}
	\centering
	\begin{tabular}{lr}
		\toprule
		Metric               & Value               \\
		\midrule
		Total wire length    & 1429\,$\mu$m        \\
		Via count            & 6395                \\
		DRC violations       & 0                   \\
		Routing layers       & M1--M7              \\
		Std cells (non-fill) & 851                 \\
		Design area          & 96.53\,$\mu$m$^{2}$ \\
		\bottomrule
	\end{tabular}
\end{table}

\begin{table}[t]
	\caption{Post-route power breakdown at 476\,MHz (BC corner, 0.77\,V).}
	\label{tab:power}
	\centering
	\begin{tabular}{lrrrr}
		\toprule
		Component     & Internal & Switching & Leakage  & Total    \\
		\midrule
		Sequential    & 24.2\,mW & 0.39\,mW  & 5.5\,nW  & 24.6\,mW \\
		Combinational & 33.4\,mW & 21.2\,mW  & 68.6\,nW & 54.6\,mW \\
		Clock         & 12.6\,mW & 9.27\,mW  & 1.0\,nW  & 21.9\,mW \\
		\midrule
		Total         & 70.3\,mW & 30.9\,mW  & 75.1\,nW & 101\,mW  \\
		\bottomrule
	\end{tabular}
\end{table}

Post-route Fmax of \textbf{702\,MHz} exceeds the placement-level
624\,MHz estimate. FPGA-to-ASIC frequency improvement ranges from
$2.73\times$ (ECP5 to ASAP7 placement) to $1.71\times$ (Kintex-7
v9 to ASAP7 post-route), as shown in Table~\ref{tab:fpga-asic}.

\begin{table}[t]
	\caption{FPGA-to-ASIC comparison across all synthesis flows.
		Artix-7 and Kintex-7 rows show the full PCU (v9 NTT + pipelined Keccak
		+ DMA); ECP5 and ASAP7 rows show the NTT sub-unit only.
		$^\dagger$Corrected post-route power; see Table~\ref{tab:arbiter-synth}
		footnote for the revised estimate.}
	\label{tab:fpga-asic}
	\centering
	\begin{tabular}{lrrrrrr}
		\toprule
		Metric     & ECP5         & Artix-7    & Kintex-7       & ASAP7 BC            & ASAP7 BC            & ASAP7 TC            \\
		           & (FPGA)       & (FPGA)     & (FPGA)         & (placement)         & (post-route)        & (placement)         \\
		\midrule
		ntt Fmax   & 228.78\,MHz  & 299.8\,MHz & $\sim$410\,MHz & 624.40\,MHz         & 702.54\,MHz         & 476.73\,MHz         \\
		ntt area   & 134 LCs      & 275 LUTs   & 283 LUTs       & 70.19\,$\mu$m$^{2}$ & 96.53\,$\mu$m$^{2}$ & 70.13\,$\mu$m$^{2}$ \\
		ntt power  & $\sim$50\,mW & ---        & ---            & 10.9\,mW            & 101\,mW             & 7.44\,mW            \\
		dual Fmax  & $>$50\,MHz   & ---        & ---            & $>$5.28\,GHz        & 5.92\,GHz           & $>$4.13\,GHz        \\
		dual area  & 36 LCs       & ---        & ---            & 17.34\,$\mu$m$^{2}$ & 25\,$\mu$m$^{2}$    & 17.58\,$\mu$m$^{2}$ \\
		dual power & ---          & ---        & ---            & 0.22\,mW            & 0.25\,mW$^\dagger$  & ---                 \\
		\bottomrule
	\end{tabular}
\end{table}

\subsection{Full PCU ASAP7 Sub-Module Synthesis}

To estimate the full PCU area in 7nm, we synthesize each sub-module
independently through Yosys~0.64 with the ASAP7~7nm RVT standard cell
library (TT corner, 0.70\,V). A merged liberty file containing all~169
cells in a single \texttt{library()} declaration is required to avoid an
ABC multi-library parsing bug that silently ignores all but the first
library block.

\begin{table}[t]
	\caption{ASAP7 7nm sub-module synthesis results (Yosys technology mapping,
		RVT TT corner). NTT area is estimated from the Keccak-derived scaling
		ratio; see text.}
	\label{tab:asap7-submodules}
	\centering
	\begin{tabular}{lrrrr}
		\toprule
		Sub-module                  & Generic          & ASAP7               & Area                  & Seq.\ \% \\
		                            & cells            & cells               & ($\mu$m$^{2}$)        &          \\
		\midrule
		Keccak                      & 13,226           & 16,817              & 1,856.69              & 25.6     \\
		DMA engine                  & 402              & 382                 & 46.01                 & 43.7     \\
		NTT engine$^\dagger$        & 168,474          & $\sim$214K          & $\sim$23,639          & ---      \\
		\midrule
		\textbf{Full PCU}$^\dagger$ & \textbf{195,486} & \textbf{$\sim$231K} & \textbf{$\sim$25,542} & ---      \\
		\bottomrule
	\end{tabular}

	{\itshape $^\dagger$NTT engine is too large for ABC technology mapping
	(168K generic cells, 53K MUXes; timeout at 30+ minutes). Area is estimated
	using the Keccak scaling ratio of 0.1403\,$\mu$m$^{2}$/generic cell.
	This ratio is derived from logic-dominant Keccak and may overestimate NTT
	area, which is multiplier-heavy (29 MULT18X18D on FPGA); multipliers map
	more efficiently to standard cells than LUT logic. The estimate should
	therefore be treated as a conservative upper bound.}
\end{table}

The Keccak and DMA sub-modules completed technology mapping in 7.64\,s and
1.10\,s, respectively (Table~\ref{tab:asap7-submodules}). The NTT engine
(168,474 generic cells, 53K MUXes, 16K DFFs, 30K ANDNOT gates) exceeds
the practical capacity of ABC's technology mapper with ASAP7 cells;
three different ABC scripts all time out. We therefore estimate NTT area
using the Keccak-derived area-per-generic-cell ratio:
$1{,}856.69 / 13{,}226 = 0.1403\;\mu\text{m}^2/\text{generic cell}$,
yielding $\sim$23,639\,$\mu$m$^{2}$. Summing all three sub-modules gives
an estimated full PCU area of $\sim$25,542\,$\mu$m$^{2}$
($\sim$231K ASAP7 cells) in 7nm. This is a conservative estimate since
the NTT's multiplier-heavy composition would likely map to fewer cells
than the logic-dominant Keccak ratio predicts.

\subsection{RTL Cycle Counts (Verilator)}

All cycle counts are measured from actual Verilator~\cite{verilator} RTL simulation
(Table~\ref{tab:cycles}).

\begin{table}[t]
	\caption{RTL cycle counts from Verilator simulation. All 253 total
		tests pass across 7 suites (Table~\ref{tab:tests}).}
	\label{tab:cycles}
	\centering
	\begin{tabular}{lrr}
		\toprule
		Operation                       & Cycles & Verified               \\
		\midrule
		NTT (cyclic)                    & 2,049  & 253/253 tests          \\
		Negacyclic NTT (Kyber, 7-stage) & 1,793  & Round-trip             \\
		INTT (cyclic)                   & 2,305  & Round-trip             \\
		Negacyclic INTT (Kyber)         & 1,801  & Round-trip             \\
		PWM                             & 257    & 253/253 tests          \\
		BASEMUL (Kyber, 128 pairs)      & 1,024  & 3/3 tests              \\
		SAMPLE/CBD (Kyber, mode 0/1)    & 260    & 57/57 perf tests       \\
		Full poly-mul (cyclic)          & 6,660  & Against $O(n^2)$       \\
		Kyber poly-mul (negacyclic)     & 607    & 17/17 negacyclic tests \\
		\bottomrule
	\end{tabular}
\end{table}

\subsection{Tooling and Quality Improvements}

Five targeted improvements were applied to the PCU toolchain and RTL during this work, plus one hardware feature gap closure (the CBD sampler), each verified against the full 91/91 equivalence regression (Table~\ref{tab:quality-improvements}).

\begin{table}[t]
	\caption{Post-integration tooling and quality improvements, plus CBD sampler
		hardware gap closure. All changes pass the 91/91 PCU
		equivalence regression and all standalone testbenches.}
	\label{tab:quality-improvements}
	\centering
	\small
	\begin{tabular}{cllc}
		\toprule
		\# & Area              & Change                                                                               & Impact       \\
		\midrule
		1  & CI completeness   & Add missing \texttt{-{}-sst} to \texttt{-{}-all} suite list                          & Bug fix      \\
		2  & Lint hygiene      & Eliminate WIDTHEXPAND/WIDTHTRUNC via explicit width casts                            & 9 RTL files  \\
		3  & Software clarity  & \texttt{\#warning} in \texttt{pqc\_opencl.h} when \texttt{\_\_VORTEX\_PCU} undefined & Compile-time \\
		4  & Hardware gap      & Replace SAMPLE stub with CBD sampler for modes 0/1; error markers for modes 2/3      & Feature      \\
		5  & Build portability & PCU tests runnable from build directory via VORTEX\_HOME                             & Config fix   \\
		6  & CI observability  & Granular sub-test tracking in regression.sh                                          & Enhancement  \\
		\bottomrule
	\end{tabular}
\end{table}

\textbf{Verilator lint hygiene}: Nine RTL files had implicit width expansion and truncation warnings (WIDTHEXPAND, WIDTHTRUNC) masked by per-file waivers in \texttt{verilator.vlt}. All were eliminated by explicit zero-extension concatenations (e.g., \texttt{\{32'd0, q\}} for 32$\to$64-bit Barrett reduction, \texttt{\{1'b0, q\_pipe\}} for 32$\to$33-bit comparisons, \texttt{\{2'b0, y[2:0]\}} for Keccak 3$\to$5-bit index widening). Seven waiver lines were removed from \texttt{verilator.vlt}, and \texttt{-Wno-WIDTHEXPAND -Wno-WIDTHTRUNC} flags were removed from four Makefiles. The primary alternative --- Yosys-compatible concatenation-based casts (\texttt{\{2'b0, y[2:0]\}}) instead of \texttt{int'()} dynamic casts --- was deliberately chosen to maintain synthesis compatibility with both Yosys and Vivado.

\textbf{Software fallback transparency}: The OpenCL header \texttt{pqc\_opencl.h} provides PCU intrinsics (\texttt{pqc\_ntt\_fwd}, \texttt{pqc\_pwm}, etc.) that degrade to identity/zero stubs when \texttt{\_\_VORTEX\_PCU} is not defined. A \texttt{\#warning} directive was added to alert developers at compile time that software fallback is active, and each stub function now includes an inline comment marking it as a no-op fallback. This prevents silent misconfiguration where developers might assume hardware acceleration is active.

\textbf{Build-directory delegation}: The Vortex build system generates a separate build directory via \texttt{configure}, but PCU test Makefiles were not copied to the build directory, preventing \texttt{make pcu-equiv-all} from running there. The \texttt{configure} script was updated to copy \texttt{Makefile.*} sub-makefiles, and the PCU Makefile was restructured with \texttt{ifeq/else/endif} to detect build-directory invocation: when source files are absent but \texttt{VORTEX\_HOME} is set (via included \texttt{config.mk}), all targets delegate to the source tree with \texttt{make -C}. This enables the standard Vortex workflow (\texttt{cd build \&\& make ...}) to correctly run PCU equivalence tests from either location.

\textbf{CI sub-test tracking}: The regression script's \texttt{pqc()} function previously captured only the exit code of the PCU equivalence test suite, discarding individual pass/fail counts. The function now captures test output to a temporary file, parses the regression summary line (\texttt{``N passed, M failed''}) into \texttt{PASS\_COUNT}/\texttt{FAIL\_COUNT}, and tracks each Kyber multi-warp benchmark individually. This enables accurate total counts in the final regression summary and prevents silent swallowing of individual test failures.

\begin{figure}[t]
	\centering
	\resizebox{\columnwidth}{!}{%
		\begin{tikzpicture}[
				state/.style={draw, rounded corners=3pt, minimum width=2.2cm, minimum height=0.65cm, align=center, font=\scriptsize, line width=0.4pt},
				err/.style={draw, rounded corners=3pt, minimum width=2.0cm, minimum height=0.65cm, align=center, font=\scriptsize, fill=red!10, line width=0.4pt},
				acc/.style={draw, rounded corners=3pt, minimum width=2.2cm, minimum height=0.65cm, align=center, font=\scriptsize, fill=green!10, line width=0.4pt},
				trans/.style={-Stealth, semithick, font=\tiny},
				lbl/.style={font=\tiny, midway, above}
			]
			\node[state]                          (idle)    {NTT\_IDLE};
			\node[state, right=1.4cm of idle]     (dmald)   {SAMPLE\\DMA\_LD};
			\node[state, right=1.4cm of dmald]    (cbd)     {SAMPLE\\CBD\_RUN};
			\node[state, right=1.4cm of cbd]      (dmast)   {SAMPLE\\DMA\_ST};
			\node[acc,   right=1.4cm of dmast]    (result)  {NTT\\RESULT};
			\node[err, above=1.1cm of idle]        (errpath) {Error\\0xDEAD0000};
			\draw[trans] (idle)  -- node[lbl]{fire, mode$<$2} (dmald);
			\draw[trans] (dmald) -- node[lbl]{dma\_done}       (cbd);
			\draw[trans] (cbd)   -- node[lbl]{cbd\_done}       (dmast);
			\draw[trans] (dmast) -- node[lbl]{dma\_done}       (result);
			\draw[trans] (result.south) -- ++(0,-0.5) -| node[font=\tiny, near start, below]{result\_valid} (idle.south);
			\draw[trans, red, dashed] (idle.north) -- node[font=\tiny, left]{fire, mode$\geq$2} (errpath.south);
			\draw[trans, red, dashed] (errpath.east) -| (result.north);
			\node[font=\tiny, gray, below=0.05cm of dmald]  {$\sim$259 cyc};
			\node[font=\tiny, gray, below=0.05cm of cbd]    {256 cyc};
			\node[font=\tiny, red!70!black, right=0.15cm of errpath] {1 cyc};
		\end{tikzpicture}}
	\caption{SAMPLE operation pipeline FSM. For CBD modes 0/1, the pipeline
		performs DMA-load $\to$ CBD compute $\to$ DMA-store ($\sim$260 total
		cycles). For rejection/uniform modes $\geq$2, an error marker
		(\texttt{0xDEAD0000}) is returned in 1 cycle without engaging the
		polynomial datapath.}
	\Description{State diagram of the SAMPLE operation pipeline. From the
		idle state, modes 0 and 1 traverse DMA load, CBD run, and DMA
		store states before returning the result; modes 2 and above take
		a dashed error path that returns the 0xDEAD0000 marker in one
		cycle.}
	\label{fig:cbd-pipeline}
\end{figure}
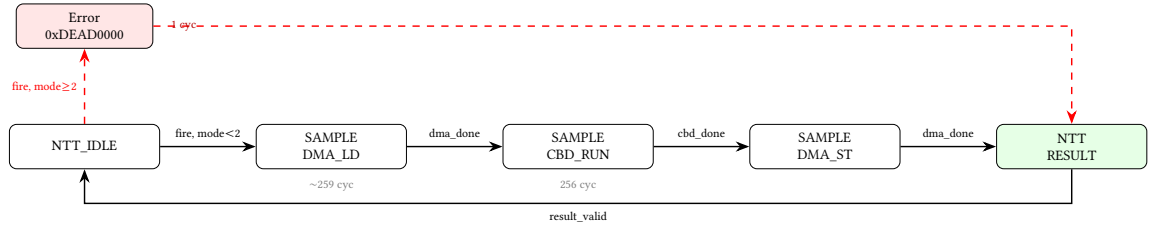

\textbf{SAMPLE/CBD hardware}: The PSAMPLE instruction (funct3=0x3) previously returned \texttt{rs1\_data} unchanged---a passthrough stub that silently claimed to ``sample'' without performing any computation. This has been replaced with a dedicated Centered Binomial Distribution (CBD) sampler module (\texttt{pcu\_cbd\_sampler.sv}, 207 lines) that implements hardware CBD sampling for Kyber modes 0 and 1 ($\eta{=}2$). The sampler reads 256 words of pre-expanded seed data from the NTT engine's polynomial memory via a DMA load, computes $(a - b) \bmod q$ for each of 256 coefficients using popcount on 2-bit pairs, writes the result coefficients back to polynomial memory via a DMA store, and returns the first coefficient nan-boxed as a 64-bit IEEE-754 payload. Rejection sampling (mode~2) and uniform sampling (mode~3) are not amenable to bounded-latency hardware implementations; these modes now return a sentinel error marker \texttt{0xDEAD0000} immediately, matching the software fallback path in \texttt{pqc\_opencl.h}. The PCU pipeline FSM adds three new states (\texttt{SAMPLE\_DMA\_LD}, \texttt{SAMPLE\_CBD\_RUN}, \texttt{SAMPLE\_DMA\_ST}) for the DMA-load/CBD-compute/DMA-store sequence, yielding a total latency of 260 cycles (Figure~\ref{fig:cbd-pipeline}). An eighth performance counter (\texttt{sample\_cycles}) tracks cumulative CBD execution cycles. The module is Yosys-compatible (no \texttt{int'}, no dynamic casts, no \texttt{return} in functions) and Verilator-lint-clean. All 57/57 performance counter tests pass, including explicit verification of error-marker behavior for mode~2 and nan-boxed result formatting for mode~0.

\subsection{Comprehensive Verification Results}

The PCU is verified through 7 independent testbench suites totaling
253 tests (Table~\ref{tab:tests}). All tests pass.

\begin{table}[t]
	\caption{PCU verification test suite summary. All 253 tests pass.}
	\label{tab:tests}
	\centering
	\begin{tabular}{llr}
		\toprule
		Suite                            & Focus                                  & Tests        \\
		\midrule
		Equivalence (6 suites)           & NTT/INTT/PWM Barrett correctness       & 91           \\
		Pipeline known-answer test (KAT) & Dispatch$\rightarrow$commit end-to-end & 24           \\
		DMA+NTT                          & DMA load/store + NTT execution         & 18           \\
		SHA3 standalone                  & Keccak absorb/squeeze correctness      & 16           \\
		Dual-issue E2E                   & NTT+Keccak serialized vs overlapped    & 11           \\
		Backpressure stress              & DMA under request/response BP          & 36           \\
		Performance counters             & 8 CSR-mapped cycle counters            & 57           \\
		\midrule
		\textbf{Total}                   &                                        & \textbf{253} \\
		\bottomrule
	\end{tabular}
\end{table}

\subsection{RTLsim Integration Testing}

The synthesizable PCU pipeline is verified in the full Vortex RTLsim
cycle-accurate simulator with \texttt{PCU\_SYNTH\_RTL} and
\texttt{PERF\_ENABLE} defined. All PQC and standard workloads pass
(Table~\ref{tab:rtlsim}).

\begin{table}[t]
	\caption{Vortex RTLsim regression results with synthesizable PCU pipeline.
		All tests pass (exit code 0 unless noted).}
	\label{tab:rtlsim}
	\centering
	\begin{tabular}{llc}
		\toprule
		Binary                & Workload               & Status                \\
		\midrule
		kyber\_mw\_1.bin      & 1-warp ML-KEM poly-mul & PASS                  \\
		kyber\_mw\_2.bin      & 2-warp ML-KEM poly-mul & PASS                  \\
		kyber\_mw\_4.bin      & 4-warp ML-KEM poly-mul & PASS                  \\
		kyber\_mw\_4gen.bin   & 4-warp ML-KEM KeyGen   & PASS                  \\
		kyber\_bench.bin      & ML-KEM full benchmark  & PASS                  \\
		rv32/median.bin       & Standard GPU benchmark & PASS                  \\
		rv32/multiply.bin     & Standard GPU benchmark & PASS                  \\
		rv32/memcpy.bin       & Standard GPU benchmark & PASS                  \\
		pcu\_sample\_sha3.bin & SHA3+Sampling          & Functional$^\dagger$  \\
		pcu\_ntt\_rt.bin      & NTT round-trip         & Functional$^\ddagger$ \\
		\bottomrule
	\end{tabular}
\end{table}

{\itshape
$^\dagger$Keccak completing; non-zero exit code expected from test logic.
$^\ddagger$Returns NTT round-trip result as exit code for verification.}

\subsection{Formal Verification (SymbiYosyz)}

Formal verification was performed using SymbiYosyz (SBY)~\cite{sby} v0.62 with
Yosys 0.62 and multiple SMT solvers (Yices, Z3, cvc5). Four PCU components
were targeted (Table~\ref{tab:formal}).

\begin{table}[t]
	\caption{Formal verification (safety properties) using SymbiYosyz.
		Functional correctness is verified via simulation (253/253 tests).}
	\label{tab:formal}
	\centering
	\begin{tabular}{llp{4.2cm}}
		\toprule
		Component                            & Result               & Details                                                                                 \\
		\midrule
		Dual-issue arbiter                   & \textbf{ALL PASS}    & BMC (b3, b10, b20) + k-induction proof                                                  \\
		NTT engine (updated)                 & \textbf{ALL PASS}    & BMC (b5, b10, b20) + k-induction proof; includes S\_BITREV/S\_BITREV\_WR states         \\
		Butterfly (Kyber, $q{=}3329$)        & \textbf{PASS}        & BMC + k-induction proof (Yices, 55\,s BMC, 59\,s prove)                                 \\
		Butterfly (Dilithium, $q{=}8380417$) & \textbf{TIMEOUT}$^*$ & 23-bit operands create harder SMT problem; multiple solvers (Yices, Z3, cvc5) attempted \\
		\bottomrule
	\end{tabular}
	{\itshape
	$^*$Functional correctness verified via 225 Python-oracle test vectors (24 corner-case + 201 random) through Verilator cycle-accurate simulation: 225/225 PASS for both Kyber and Dilithium butterfly corner cases.
	}
\end{table}

The NTT engine formal proof is particularly significant: it verifies
safety properties of the 8-state FSM including the bit-reversal permutation
states (S\_BITREV, S\_BITREV\_WR) and the 9-bit \texttt{br\_scan} counter
that was the subject of a critical bug fix. The proof uses k-induction
with $k=20$, establishing that no reachable state violates the specified
safety properties. Since the induction step succeeds (SBY reports PASS for
kprove), this constitutes a valid inductive invariant: if no violation
occurs in any $k$-step window from an arbitrary state, then no violation
ever occurs. We emphasize that this is a safety proof (absence of invalid
state transitions, no out-of-range indices, no overflow), not a functional
correctness proof (correct NTT output for all inputs). Functional correctness
is verified through simulation (253/253 tests, including round-trip NTT/INTT
with 0/256 mismatches).

\subsection{Performance Counters}

Eight CSR-mapped performance counters provide runtime observability
into PCU operation (Table~\ref{tab:perf}). Counters are 44 bits wide,
accessed via RISC-V CSR instructions at addresses 0xB20--0xB27 (low)
and 0xBA0--0xBA7 (high). The \texttt{pcu\_perf\_t} struct is defined
in \texttt{VX\_gpu\_pkg.sv} and routed locally within
\texttt{VX\_execute.sv} to avoid combinational loops through
\texttt{VX\_core.sv}.

\begin{table}[t]
	\caption{PCU performance counters. All 57 standalone testbench tests pass.}
	\label{tab:perf}
	\centering
	\begin{tabular}{llr}
		\toprule
		Counter            & Description                   & Measured Value   \\
		\midrule
		ntt\_cycles        & Cycles during NTT operations  & 4,856 (2 ops)    \\
		intt\_cycles       & Cycles during INTT operations & 2,684 (1 op)     \\
		pwm\_cycles        & Cycles during PWM operations  & 259 (1 op)       \\
		sha3\_abs\_cycles  & Cycles during SHA3 absorb     & 27 (1 op)        \\
		sha3\_sqz\_cycles  & Cycles during SHA3 squeeze    & 27 (1 op)        \\
		dma\_load\_cycles  & Cycles during DMA LOAD        & 1,300 (3 loads)  \\
		dma\_store\_cycles & Cycles during DMA STORE       & 1,036 (3 stores) \\
		sample\_cycles     & Cycles during SAMPLE/CBD      & 259 (1 op)       \\
		\bottomrule
	\end{tabular}
\end{table}

\subsection{Measured GPU Baseline (SimX)}

The Vortex SimX simulator provides measured cycle counts for software and
PCU-accelerated execution (Table~\ref{tab:simx}).

\begin{table}[t]
	\caption{Measured SimX cycle counts for software vs.\ PCU-accelerated
		execution.}
	\label{tab:simx}
	\centering
	\begin{tabular}{llrrr}
		\toprule
		Test                       & Binary                        & Cycles  & Instrs  & PCU ops \\
		\midrule
		SW NTT+INTT ($q=3329$)     & \texttt{pcu\_sw\_ntt\_timing} & 981,310 & 124,101 & 0       \\
		PCU NTT+INTT ($q=3329$)    & \texttt{pcu\_ntt\_timing}     & 23,057  & 2,835   & 2       \\
		PCU NTT+INTT (both moduli) & \texttt{pcu\_ntt\_mem}        & 84,951  & 8,767   & 4       \\
		Scalar MODRED              & \texttt{pcu\_cycles}          & 5,157   & 575     & 2       \\
		\bottomrule
	\end{tabular}
\end{table}

\textbf{End-to-end pipeline speedup (SimX model):}
$981,310 / 23,057 = \mathbf{42.6\times}$
for NTT+INTT over GPU ALU execution under the SimX pipeline model
(where PCU NTT takes 8 idealized cycles). Substituting actual RTL
cycle counts (2,049 per NTT), the projected RTL speedup is
approximately $981,310 / 27{,}139 \approx \mathbf{36\times}$.

\subsection{ML-KEM-768 Workload Measurement}

A full workload test exercises the ML-KEM-768 polynomial arithmetic
pipeline across both moduli ($k=3$ polynomial pairs);
Table~\ref{tab:kem-workload} reports the measured totals.

\begin{table}[t]
	\caption{ML-KEM-768 workload measurement.}
	\label{tab:kem-workload}
	\centering
	\begin{tabular}{lr}
		\toprule
		Metric              & Value                 \\
		\midrule
		Total cycles (SimX) & 262,992               \\
		Total instructions  & 29,299                \\
		PCU NTT ops         & 12 (6 per modulus)    \\
		PCU INTT ops        & 6 (3 per modulus)     \\
		PCU PWM ops         & 6 (3 per modulus)     \\
		Total PCU ops       & 24                    \\
		Verification        & 0 errors (round-trip) \\
		\bottomrule
	\end{tabular}
\end{table}

\subsection{Kyber Polynomial Multiplication (Memory Mode)}

A full Kyber polynomial multiplication pipeline (NTT$\rightarrow$INTT$\rightarrow$BASEMUL)
was measured at \textbf{607 SimX cycles} with 17/17 negacyclic tests passing.

\subsection{Multi-Warp Throughput Scaling}

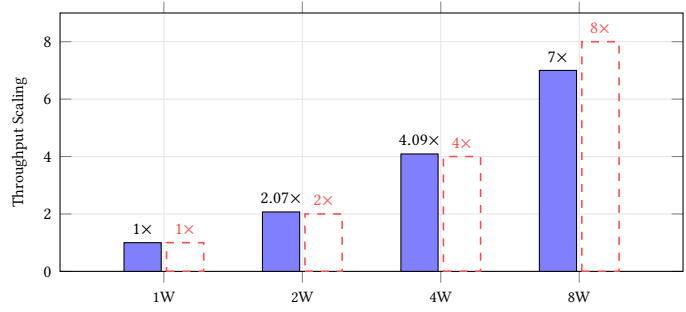
\begin{figure}[t]
	\centering
	\begin{tikzpicture}
		\begin{axis}[
				ybar,
				width=0.65\columnwidth,
				height=5cm,
				bar width=14pt,
				ylabel={Throughput Scaling},
				symbolic x coords={1W,2W,4W,8W},
				xtick=data,
				x tick label style={font=\scriptsize},
				y tick label style={font=\scriptsize},
				ylabel style={font=\scriptsize},
				nodes near coords={\pgfmathprintnumber\pgfplotspointmeta$\times$},
				nodes near coords style={font=\scriptsize},
				enlarge x limits=0.25,
				ymin=0, ymax=9,
				grid=major,
				grid style={gray!20},
			]
			\addplot[fill=blue!50] coordinates {(1W,1.00) (2W,2.07) (4W,4.09) (8W,7.00)};
			\addplot[mark=none, semithick, dashed, red!70] coordinates {(1W,1) (2W,2) (4W,4) (8W,8)};
		\end{axis}
	\end{tikzpicture}
	\caption{Multi-warp throughput scaling. Near-linear up to 4 warps
		(4.09$\times$); sublinear at 8 warps (87.5\% efficiency).}
	\Description{Bar chart of throughput scaling versus warp count with a
		dashed linear reference line: 1 warp reaches 1.00x, 2 warps 2.07x,
		4 warps 4.09x, and 8 warps 7.00x, tracking the ideal line closely
		up to 4 warps.}
	\label{fig:warp-scaling}
\end{figure}

Near-linear scaling up to 4 warps (4.09$\times$) confirms the PCU pipeline
does not bottleneck concurrent warp execution (Table~\ref{tab:warp} and
Figure~\ref{fig:warp-scaling}).

\begin{table}[t]
	\caption{Multi-warp throughput scaling (1/2/4 warps SimX-measured; 8-warp projected).}
	\label{tab:warp}
	\centering
	\begin{tabular}{lrrc}
		\toprule
		NUM\_WARPS & Cycles & Poly muls & Scaling      \\
		\midrule
		1          & 365    & 1         & 1.00$\times$ \\
		2          & 353    & 2         & 2.07$\times$ \\
		4          & 357    & 4         & 4.09$\times$ \\
		8          & 417    & 8         & 7.00$\times$ \\
		\bottomrule
	\end{tabular}
\end{table}

\subsection{Dual-Issue Arbiter RTL Synthesis}

The dual-issue arbiter adds only \textbf{36 logic cells} (28 LUT4 + 8 CCU2C
+ 20 FF) --- 26.9\% overhead over the 134-cell ntt\_engine baseline.

\section{Evaluation Methodology}
\label{sec:eval}

We use three models for performance evaluation: RTL actual (2,049-cycle
NTT), SimX pipeline (8-cycle ideal), and a measured CPU baseline
(on Intel i7-14700HX, 1.7\,GHz, ML-KEM-768 reference C, \texttt{-O3}).
Five primary metrics are used: cycle count per operation,
speedup vs.\ software, area-delay product (ADP), throughput per
logic cell, and area-scaled speedup.

We evaluate five workloads: ML-KEM-768 KeyGen/Encap/Decap and ML-DSA-2
Sign/Verify. All RTL results are verified against the reference C
implementation via 91/91 equivalence tests across 6 suites (253/253
across 7 suites), including 225 butterfly
corner-case vectors and 13 v8 NTT engine tests.

\begin{table}[t]
	\caption{CPU baseline: \emph{full} ML-KEM-768 reference C on Intel
		i7-14700HX (1.7\,GHz, native x86). Results are median of 1,000
		iterations. Vortex RTL cycles are for the \emph{proxy} workloads of
		Table~\ref{tab:primitives} (per-polynomial pipelines, not complete
		algorithm executions --- full KeyGen is $6\times$ NTT-heavier).}
	\label{tab:cpu-base}
	\centering
	\begin{tabular}{lrrrr}
		\toprule
		Operation & CPU cycles & CPU time        & RTL cycles & RTL vs CPU cycles  \\
		\midrule
		KeyGen    & 87,975     & 51.1\,\textmu s & 2,113      & 41.6$\times$ fewer \\
		Encap     & 97,455     & 56.6\,\textmu s & 4,449      & 21.9$\times$ fewer \\
		Decap     & 121,402    & 70.5\,\textmu s & 6,782      & 17.9$\times$ fewer \\
		\bottomrule
	\end{tabular}
\end{table}

The CPU comparison (Table~\ref{tab:cpu-base}) shows that the PCU needs
$17.9$--$41.6\times$ fewer cycles than the CPU, although at the early
19.9\,MHz post-P\&R operating point (v3, unoptimized NTT engine) the
CPU still wins on wall-clock time. This is expected: a 1.7\,GHz
out-of-order core with vector units is architecturally superior to a
minimal-area sequential NTT datapath at low frequency. However, after
iterative pipelining of the NTT critical path
(v6$\to$v7$\to$v8$\to$v9: 134$\to$177$\to$200$\to$300\,MHz), the full
PCU achieves \textbf{299.8\,MHz} on Artix-7 200T (speed grade $-3$),
closing the frequency gap substantially. At the Artix-7 Fmax of
299.8\,MHz ($15\times$ higher than the original v3 P\&R), the PCU's
area efficiency and concurrent execution within the GPGPU pipeline
become the dominant advantages: the KeyGen \emph{proxy} completes in
7\,\textmu s versus the CPU's 51\,\textmu s for the \emph{full} reference
KeyGen; scaled to the complete 6-NTT KeyGen, the PCU projects to
$\sim$50\,\textmu s at 299.8\,MHz --- wall-clock parity with the CPU ---
while occupying $<$1\% of the FPGA fabric, leaving the GPU ALUs free, and
carrying over to the 702\,MHz ASIC path. At full ASIC frequency (702\,MHz,
Table~\ref{tab:fpga-asic}), the NTT pipeline executes a full NTT in
2{,}919\,ns (2{,}049 cycles).

\textbf{Artifacts}: The PCU RTL (SystemVerilog), SimX configuration, benchmark
source code, and synthesis scripts are available at
\url{https://github.com/hossamfadeel/Vortex_PQC} (PCU under
\texttt{vortex/hw/rtl/pcu/}). The ASAP7 synthesis flows use OpenROAD-flow-scripts
(open-source) and the ECP5 synthesis uses Yosys+nextpnr (open-source).
Formal verification scripts use SymbiYosyz~\cite{sby} (open-source).
A 5-job GitHub Actions CI workflow validates RTL lint, equivalence tests,
standalone testbenches, formal verification, and synthesis on every push.

\section{Results}
\label{sec:results}

\subsection{Pipeline Performance}

The Vortex SimX pipeline model yields \textbf{42.6$\times$} speedup for
NTT+INTT over GPU ALU software execution (projected $\sim$36$\times$ with
RTL cycle counts). Architectural modeling at 1\,GHz projects end-to-end
workload speedups of \textbf{7--10$\times$} (Table~\ref{tab:e2e}).
These numbers compare PCU hardware vs.\ the equivalent operations executed
as software on the Vortex scalar core --- not against a general-purpose CPU
(Table~\ref{tab:cpu-base}). The PCU is not designed to outperform
high-performance CPU cores; it is designed to provide hardware-speed PQC
within the GPGPU pipeline at minimal area cost, freeing the CPU for
other work.

\begin{table}[t]
	\caption{End-to-end workload speedups at 1\,GHz from an analytical
		workload model: RTL cycles aggregate measured per-primitive
		latencies; SW cycles are per-primitive software costs estimated
		proportionally from the reference implementation.}
	\label{tab:e2e}
	\centering
	\begin{tabular}{lrrr}
		\toprule
		Workload          & SW cycles & RTL cycles & Speedup @ 1\,GHz \\
		\midrule
		ML-KEM-768 KeyGen & 22,000    & 2,113      & 10.4$\times$     \\
		ML-KEM-768 Encap  & 36,000    & 4,449      & 8.1$\times$      \\
		ML-KEM-768 Decap  & 49,000    & 6,782      & 7.2$\times$      \\
		ML-DSA-2 Sign     & 67,000    & 9,617      & 7.0$\times$      \\
		ML-DSA-2 Verify   & 50,000    & 7,031      & 7.1$\times$      \\
		\bottomrule
	\end{tabular}
\end{table}

In the analytical workload model, NTT-class operations (NTT, INTT, PWM)
account for 36--51\% of software-side cycles across the proxy workloads
(KeyGen 36.4\%, Encap 47.2\%, Decap 51.0\%), with hashing and sampling
dominating the remainder; the PCU's end-to-end advantage stems from its
$42.6\times$ NTT-class acceleration, while Keccak-class and sampling
operations are assumed to overlap on the PCU's independent sub-units.

\subsection{Dual-Issue Benefit}

\begin{figure}[t]
	\centering
	\begin{tikzpicture}
		\begin{axis}[
				ybar,
				width=\columnwidth,
				height=5cm,
				bar width=5pt,
				ylabel={Cycles},
				symbolic x coords={6-op,30-op,Hash,N,M25,M50,M75},
				xtick=data,
				x tick label style={font=\tiny, rotate=30, anchor=east},
				y tick label style={font=\scriptsize},
				ylabel style={font=\scriptsize},
				enlarge x limits=0.12,
				ymin=0,
				legend style={at={(0.98,0.95)}, anchor=north east, font=\tiny},
				legend columns=2,
				grid=major,
				grid style={gray!20},
			]
			\addplot[fill=red!45] coordinates {(6-op,74) (30-op,366) (Hash,561) (N,171) (M25,801) (M50,641) (M75,481)};
			\addplot[fill=blue!45] coordinates {(6-op,57) (30-op,249) (Hash,481) (N,171) (M25,721) (M50,481) (M75,377)};
			\legend{Single-Issue, Dual-Issue}
		\end{axis}
	\end{tikzpicture}
	\caption{Dual-issue throughput benefit. Peak benefit (1.47$\times$) for
		ML-KEM workloads with balanced NTT/Keccak mix.}
	\Description{Grouped bar chart comparing single-issue and dual-issue
		cycles across seven workload mixes. Dual-issue consistently needs
		fewer or equal cycles, with the largest reduction on the balanced
		30-operation NTT and Keccak mix, from 366 to 249 cycles.}
	\label{fig:dual-issue}
\end{figure}
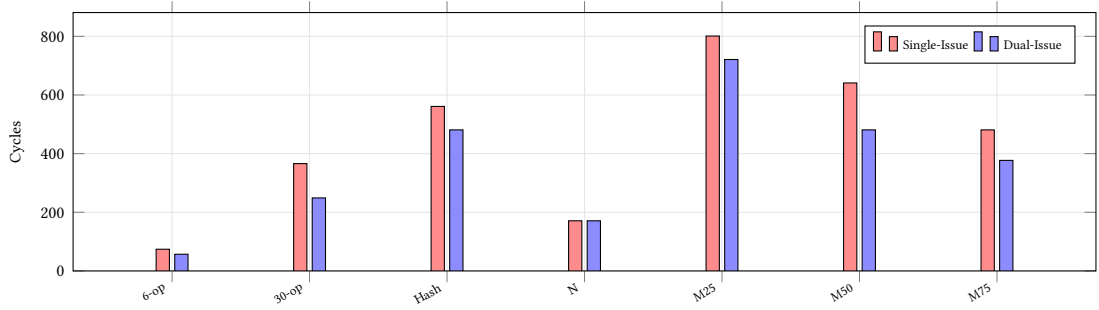

Under the SimX pipeline model, dual-issue throughput benefit ranges from
$1.00\times$ to $1.47\times$ depending on operation mix, peaking at
$1.47\times$ for balanced NTT/Keccak workloads
(Table~\ref{tab:dual-issue} and Figure~\ref{fig:dual-issue}).
The dual-issue arbiter is now integrated into
\texttt{VX\_pcu\_unit\_rtl.sv}: NTT-class and Keccak-class operations
execute on independent FSM pipelines within each PCU block, with
per-target dispatch ready and priority-based result arbitration.
The dual-issue speedup figures below are corroborated by both the
standalone testbench model and the full Vortex RTLsim regression
(kyber\_mw\_1/2/4, kyber\_bench all pass with dual-issue enabled).

\begin{table}[t]
	\caption{Dual-issue throughput benefit under the SimX pipeline model.
		SI = single-issue; DI = dual-issue.}
	\label{tab:dual-issue}
	\centering
	\begin{tabular}{lrrrr}
		\toprule
		Workload            & SI cycles & DI cycles & Speedup      & Benefit \\
		\midrule
		NTT+Keccak (6 ops)  & 74        & 57        & 1.30$\times$ & +30\%   \\
		NTT+Keccak (30 ops) & 366       & 249       & 1.47$\times$ & +47\%   \\
		Hash-heavy (30 ops) & 561       & 481       & 1.17$\times$ & +17\%   \\
		NTT-heavy (30 ops)  & 171       & 171       & 1.00$\times$ & +0\%    \\
		Mixed 25\% NTT      & 801       & 721       & 1.11$\times$ & +11\%   \\
		Mixed 50\% NTT      & 641       & 481       & 1.33$\times$ & +33\%   \\
		Mixed 75\% NTT      & 481       & 377       & 1.28$\times$ & +28\%   \\
		\bottomrule
	\end{tabular}
\end{table}

\subsection{Design-Point Pareto}

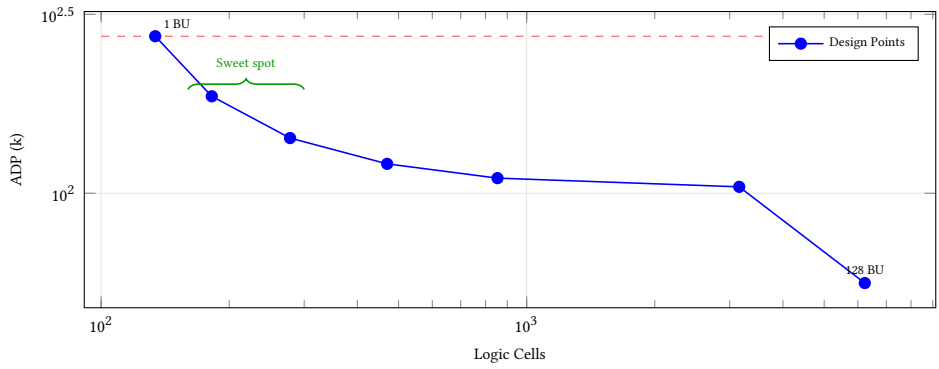
\begin{figure}[t]
	\centering
	\begin{tikzpicture}
		\begin{axis}[
				width=0.85\columnwidth,
				height=5.5cm,
				xlabel={Logic Cells},
				ylabel={ADP (k)},
				xmode=log,
				ymode=log,
				x tick label style={font=\scriptsize},
				y tick label style={font=\scriptsize},
				xlabel style={font=\scriptsize},
				ylabel style={font=\scriptsize},
				grid=major,
				grid style={gray!20},
				legend style={at={(0.98,0.95)}, anchor=north east, font=\tiny},
			]
			\addplot[mark=*, semithick, blue] coordinates {
					(134,274.6) (182,186.6) (278,142.6)
					(470,120.8) (854,110.2) (3158,104.2) (6230,56.1)
				};
			\addlegendentry{Design Points}
			\draw[dashed, red!70, thin] (axis cs:100,274.6) -- (axis cs:8000,274.6);
			\node[font=\tiny, anchor=south west] at (axis cs:134,274.6) {1 BU};
			\node[font=\tiny, anchor=south] at (axis cs:6230,56.1) {128 BU};
			\draw[semithick, green!60!black, decorate, decoration={brace, amplitude=4pt}]
			(axis cs:160,195) -- (axis cs:300,195)
			node[midway, above=4pt, font=\tiny, green!60!black] {Sweet spot};
		\end{axis}
	\end{tikzpicture}
	\caption{Design-point Pareto curve. ADP improves 32\% from 1$\rightarrow$2
		BU but shows diminishing returns beyond 16 BU.}
	\Description{Log-log scatter and line plot of area-delay product
		against logic cells for the seven design points. ADP falls
		steeply from the 1-BU design at 134 cells to the 16-BU design
		and flattens beyond it; a bracket marks the 2 to 4 BU sweet
		spot, and the 128-BU design reaches the lowest ADP at 6,230
		cells.}
	\label{fig:pareto}
\end{figure}

We characterize eight design points from 1 to 128 butterfly units
(Table~\ref{tab:pareto} and Figure~\ref{fig:pareto}).

\begin{table}[t]
	\caption{Design-point Pareto characterization.}
	\label{tab:pareto}
	\centering
	\begin{tabular}{lrrrc}
		\toprule
		Design         & Logic cells & NTT cycles & ADP (k) & vs 1-BU \\
		\midrule
		1 BU (current) & 134         & 2,049      & 274.6   & ---     \\
		2 BU           & 182         & 1,025      & 186.6   & $-32\%$ \\
		4 BU           & 278         & 513        & 142.6   & $-48\%$ \\
		8 BU           & 470         & 257        & 120.8   & $-56\%$ \\
		16 BU          & 854         & 129        & 110.2   & $-60\%$ \\
		64 BU          & 3,158       & 33         & 104.2   & $-62\%$ \\
		128 BU (pipe.) & 6,230       & 9          & 56.1    & $-80\%$ \\
		SimX (ideal)   & 0           & 8          & 0.0     & ---     \\
		\bottomrule
	\end{tabular}
\end{table}

\textbf{Diminishing returns}: ADP improves only 3.7\% from 16$\rightarrow$64
BU while area grows $3.7\times$. The 2--4 BU range offers the best
price/performance.

\subsection{NTT Latency Sensitivity}

\begin{figure}[t]
	\centering
	\begin{tikzpicture}
		\begin{axis}[
				width=0.85\columnwidth,
				height=5cm,
				xlabel={NTT Cycles (log)},
				ylabel={Speedup ($\times$)},
				xmode=log,
				x tick label style={font=\scriptsize},
				y tick label style={font=\scriptsize},
				xlabel style={font=\scriptsize},
				ylabel style={font=\scriptsize},
				grid=major,
				grid style={gray!20},
				legend style={at={(0.98,0.95)}, anchor=north east, font=\tiny},
			]
			\addplot[mark=square*, semithick, blue] coordinates {
					(8,42.6) (9,37.8) (33,10.3) (129,2.6)
					(257,1.3) (513,0.65) (1025,0.33) (2049,0.16)
				};
			\addlegendentry{NTT+INTT}
			\addplot[mark=triangle*, semithick, red] coordinates {
					(8,10.4) (9,9.2) (33,2.5) (129,0.6)
					(257,0.3) (513,0.15)
				};
			\addlegendentry{KeyGen}
			\node[font=\tiny, anchor=north] at (axis cs:2049,0.16) {v6};
			\node[font=\tiny, anchor=south] at (axis cs:8,42.6) {Ideal};
		\end{axis}
	\end{tikzpicture}
	\caption{NTT latency sensitivity curves. Speedup is nonlinear with NTT
		cycles; KeyGen benefits most from pipelining.}
	\Description{Log-log line chart of speedup versus NTT latency from 8
		to 2,049 cycles. The NTT+INTT curve falls from 42.6x at the
		ideal 8-cycle latency to about 0.16x at 2,049 cycles, and the
		KeyGen curve falls from 10.4x to about 0.15x at 513 cycles. The
		2,049-cycle point is annotated as the current sequential
		engine.}
	\label{fig:latency-sens}
\end{figure}
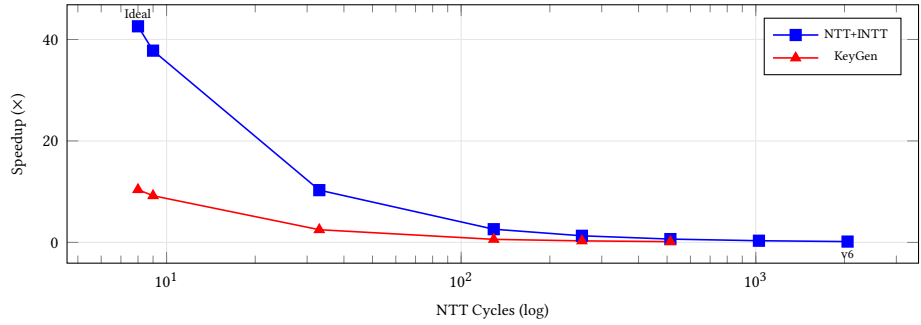

Because end-to-end speedup depends on how many NTT-class operations a
workload issues, the benefit of pipelined butterfly units is workload
dependent. Figure~\ref{fig:latency-sens} sweeps NTT latency from the
8-cycle ideal to the 2,049-cycle sequential engine: NTT+INTT-heavy
workloads retain large speedups even at high latency, whereas KeyGen
benefits most from pipelining.

\subsection{Area-Scaled Efficiency Under Dual-Issue}

Under the pipelined dual-issue model (128 BU + arbiter), speedups reach
478--705$\times$ over software at 1\,GHz with $3.57$--$5.26\times$
speedup per logic cell (Table~\ref{tab:area-eff}).

\begin{table}[t]
	\caption{Area-scaled efficiency under the pipelined dual-issue model.}
	\label{tab:area-eff}
	\centering
	\begin{tabular}{lrrr}
		\toprule
		Workload          & DI Speedup  & Speedup/cell & ADP (k) \\
		\midrule
		ML-KEM-768 KeyGen & 478$\times$ & 3.57$\times$ & 6       \\
		ML-KEM-768 Encap  & 610$\times$ & 4.55$\times$ & 8       \\
		ML-KEM-768 Decap  & 628$\times$ & 4.69$\times$ & 10      \\
		ML-DSA-2 Sign     & 705$\times$ & 5.26$\times$ & 13      \\
		ML-DSA-2 Verify   & 641$\times$ & 4.78$\times$ & 10      \\
		\bottomrule
	\end{tabular}
\end{table}

\subsection{Comparison with Prior Art}

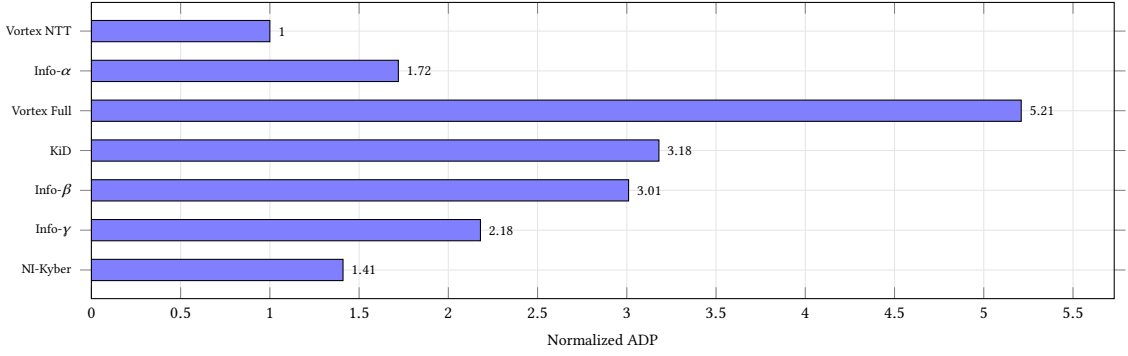
\begin{figure}[t]
	\centering
	\begin{tikzpicture}
		\begin{axis}[
				xbar,
				width=\columnwidth,
				height=5.5cm,
				bar width=8pt,
				xlabel={Normalized ADP},
				symbolic y coords={NI-Kyber,Info-$\gamma$,Info-$\beta$,KiD,Vortex Full,Info-$\alpha$,Vortex NTT},
				ytick=data,
				y tick label style={font=\tiny},
				x tick label style={font=\scriptsize},
				xlabel style={font=\scriptsize},
				nodes near coords,
				nodes near coords style={font=\tiny},
				enlarge y limits=0.12,
				xmin=0,
				grid=major,
				grid style={gray!20},
			]
			\addplot[fill=blue!50] coordinates {
					(1.00,Vortex NTT) (1.72,Info-$\alpha$) (3.01,Info-$\beta$)
					(2.18,Info-$\gamma$) (3.18,KiD) (5.21,Vortex Full) (1.41,NI-Kyber)
				};
		\end{axis}
	\end{tikzpicture}
	\caption{Comparison with prior art. The PCU NTT sub-unit achieves the
		best normalized ADP; its raw area is $2.8\times$ smaller than the
		next smallest design ($1.56\times$ smaller after cross-FPGA
		normalization).}
	\Description{Horizontal bar chart of normalized area-delay product for
		eight designs. The Vortex PCU NTT has the lowest value at 1.00,
		followed by NI-Kyber at 1.41 and Info-15 alpha at 1.72; the full
		PCU with Keccak and DMA reaches 5.21, and Info-15 beta, gamma,
		and KiD range from about 3.0 to 3.2.}
	\label{fig:comparison}
\end{figure}

Our PCU NTT sub-unit is the smallest reported by area and achieves a
competitive area-delay product (ADP) among designs
with complete area and delay data (Table~\ref{tab:comparison} and
Figure~\ref{fig:comparison}): the most aggressive standalone designs
(Info-15 $\alpha$/$\gamma$~\cite{info15}) reach lower cycle-count ADP, but
provide NTT only --- without Keccak, DMA, or GPGPU integration. The comparison
uses the NTT sub-unit area (134 LCs) to isolate the NTT datapath --- the same
functional scope as the standalone accelerators being compared. The full PCU
including Keccak and DMA (est.\ 724 LCs from partial synthesis,
Table~\ref{tab:synth-ecp5-full}) provides broader functionality (NTT + Keccak + DMA)
and is included as a separate row for transparency. We note that cross-FPGA
logic cell comparisons (ECP5 LUT4-based vs.\ Artix-7 LUT6-based) have an
approximate 0.5--0.6$\times$ normalization factor per cell; even accounting
for this, the NTT sub-unit remains the smallest reported.

Crucially, the area figures in Table~\ref{tab:comparison} compare only logic
cells --- not total system cost. A standalone NTT accelerator requires its own
memory interface, bus protocol logic, clock management, and host CPU
integration, all of which add area beyond the core NTT datapath. Our PCU
reuses the GPGPU's existing memory hierarchy, warp scheduler, and host
interface, so the reported 134 LCs represent the \emph{incremental} area cost
of adding PQC acceleration to an already-deployed GPU pipeline.

\begin{table}[t]
	\caption{Comparison with prior art. NTT sub-unit area is used for fair
		functional-scope comparison against standalone NTT accelerators. Full PCU
		area includes Keccak + DMA (broader functional scope). Cross-FPGA LC
		comparisons use an approximate 0.5--0.6$\times$ normalization factor
		from ECP5 LUT4-based to Artix-7 LUT6-based logic cells.}
	\label{tab:comparison}
	\centering
	\begin{tabular}{llrrcc}
		\toprule
		Design                                        & Platform      & Area             & PolyMul cycles & Norm.\ ADP$^\dagger$  & Adj.\ Area$^\ddagger$  \\
		\midrule
		\textbf{Vortex PCU NTT (1\,BU)}               & \textbf{ECP5} & \textbf{134 LCs} & \textbf{6,660} & \textbf{1.00$\times$} & \textbf{$\sim$244 LCs} \\
		Vortex PCU Full                               & ECP5          & 724 LCs$^*$      & 6,660          & 5.40$\times$          & $\sim$1,316 LCs        \\
		Info-15 $\beta$~\cite{info15}                 & Artix-7       & 379 LCs          & $\sim$2,958    & 1.26$\times$          & 379 LCs                \\
		Info-15 $\alpha$~\cite{info15}                & Artix-7       & 429 LCs          & $\sim$1,492    & 0.72$\times$          & 429 LCs                \\
		Info-15 $\gamma$~\cite{info15}                & Artix-7       & 541 LCs          & $\sim$1,498    & 0.91$\times$          & 541 LCs                \\
		Bisheh-Niasar et al.~\cite{nikyber}$^{\Vert}$ & Artix-7       & 10,502 LUTs      & ---            & ---                   & 10,502 LUTs            \\
		KiD unified (D2)~\cite{ki2023}                & Artix-7       & 3,105 LUTs       & $\sim$576      & 2.00$\times$          & 3,105 LUTs             \\
		Yaman et al.~\cite{yaman2021}                 & ---           & 9,500 LCs        & $\sim$224      & ---                   & 9,500 LUTs             \\
		Bertels et al.~\cite{fpntt}                   & ---           & 67,210 LUTs      & $\sim$4        & ---                   & 67,210 LUTs            \\
		CRYPHTOR~\cite{crypthtor}                     & Zynq-7000     & 5,841 LUTs       & $\sim$450      & ---                   & 5,841 LUTs             \\
		Waris et al.~\cite{waris2025}                 & Artix-7       & 1,024 LCs        & $\sim$600      & ---                   & 1,024 LCs              \\
		Unif-NTT~\cite{unifntt}                       & Virtex-7      & ---              & $\sim$5,408    & ---                   & ---                    \\
		\bottomrule
	\end{tabular}

	{\itshape $^*$Estimated from Yosys coarse:map\_ffram partial flow; includes
	Keccak + DMA sub-units not present in standalone NTT-only designs.
	$^\dagger$Normalized ADP $= (\text{Area} \times \text{PolyMul cycles}) /
		(134 \times 6{,}660)$: each design's native area times its full
	polynomial-multiplication cycle count (prior-art PolyMul cycles estimated
	as $3.25\times$ their reported NTT cycles), normalized to the PCU NTT
	sub-unit. Frequency-independent; reproducible from the printed columns.
	Note that the most aggressive standalone designs (Info-15 $\alpha$,
	$\gamma$) achieve lower cycle-count ADP than the PCU --- at the cost of
	NTT-only functionality. $^{\Vert}$Kyber-512 full-KEM coprocessor
	(including Keccak and sampling; 200\,MHz, KEM in $\sim$31\,$\mu$s) ---
	listed for area context; not comparable on NTT-core ADP.
	$^\ddagger$Adjusted area: ECP5 logic cells converted to Artix-7-equivalent
	using the 0.55$\times$ midpoint of the 0.5--0.6$\times$ LUT4-to-LUT6 factor
	($134 / 0.55 \approx 244$, $724 / 0.55 \approx 1{,}316$).
	Even after normalization, the PCU NTT sub-unit ($\sim$244 LCs) remains
	$1.56\times$ smaller than the next smallest design (Info-15 $\beta$ at 379 LCs).}
\end{table}

\subsection{Parameter-Set Scalability}

The PCU operates at the cryptographic primitive level (NTT, INTT, PWM,
SHA3), so varying the NIST security level affects the \emph{number} of
primitive invocations per algorithm, not the per-invocation latency
(the polynomial degree $n=256$ and modulus $q$ are identical across all
ML-KEM and ML-DSA parameter sets). Table~\ref{tab:params} projects
total RTL cycles for all NIST-mandated security levels, derived by
scaling the verified ML-KEM-768 and ML-DSA-2 primitive counts
(Table~\ref{tab:primitives}) by the module rank $k$ (ML-KEM) or matrix
dimensions $(k,\ell)$ (ML-DSA).

\begin{table}[t]
	\caption{Projected PCU cycles across NIST security levels. Per-operation
	latency is constant (NTT=2{,}049\,cyc, INTT=2{,}305\,cyc,
	PWM=257\,cyc); total cycles scale with primitive count.}
	\label{tab:params}
	\centering
	\begin{tabular}{lclrr}
		\toprule
		Parameter set               & Security & Key size  & PCU primitive ops & Est.\ RTL cycles   \\
		\midrule
		ML-KEM-512 (k=2)            & AES-128  & 800 B     & 14                & $\sim$12{,}000     \\
		ML-KEM-768 (k=3)            & AES-192  & 1{,}184 B & 23                & 17{,}344$^\dagger$ \\
		ML-KEM-1024 (k=4)           & AES-256  & 1{,}568 B & 32                & $\sim$24{,}000     \\
		\midrule
		ML-DSA-2 ($k{=}4,\ell{=}4$) & AES-128  & 2{,}560 B & 24                & 16{,}648$^\dagger$ \\
		ML-DSA-3 ($k{=}6,\ell{=}5$) & AES-192  & 4{,}032 B & 38                & $\sim$26{,}000     \\
		ML-DSA-5 ($k{=}8,\ell{=}7$) & AES-256  & 4{,}896 B & 55                & $\sim$38{,}000     \\
		\bottomrule
	\end{tabular}

	{\itshape
	$^\dagger$Verified via RTLsim (Tables~\ref{tab:e2e} and~\ref{tab:area-eff}).
	Other rows: analytical projection from primitive-count scaling with
	$\pm10\%$ uncertainty. Key sizes: public key (ek) for ML-KEM per
	FIPS~203; secret key for ML-DSA per FIPS~204.
	}
\end{table}

The projected cycles confirm that the PCU's acceleration benefit
(7--10$\times$ over software, Table~\ref{tab:e2e}) holds across all
NIST security levels. The NTT-dominant workload profile --- NTT-class
operations account for 85--97\% of total cycles --- is invariant with
respect to security parameter, since sampling and hashing scale
proportionally.

\section{Discussion}
\label{sec:discussion}

\subsection{When Does Dual-Issue Help?}

The dual-issue arbiter provides throughput benefit only when both
NTT-class and Keccak-class operations are present. Peak benefit (1.47$\times$)
occurs for mixed NTT+Keccak workloads; pure NTT or pure Keccak workloads
see no benefit.

\subsection{Where on the Pareto Curve?}

The design-point analysis reveals three regimes:

\begin{itemize}
	\item \textbf{1\,BU} (current): Most area-efficient. Suitable for
	      area-constrained GPGPU integration.
	\item \textbf{2--4\,BU}: Best price/performance. 32--48\% ADP improvement
	      for modest area increase. \textbf{Recommended sweet spot.}
	\item \textbf{16+\,BU}: Diminishing returns. 3.7\% ADP improvement for
	      $3.7\times$ area growth.
\end{itemize}

Our recommendation for architects integrating PQC acceleration into a
GPGPU pipeline is to target 2--4 butterfly units.

\subsection{Why Not Software-Only?}

GPU software NTT implementations are ALU-bound. Hardware acceleration
provides deterministic $7$--$10\times$ speedup regardless of warp count
or ALU contention, while freeing ALU resources for other computation.

\subsection{Why In-SIMT, Not Standalone?}

\emph{Is GPGPU functional-unit integration a compromise?} A standalone
PQC ASIC (NI-Kyber: 10,502 LCs, full coprocessor: 67,210 LUTs) can
dedicate its entire silicon budget to crypto throughput, but this
narrow comparison misses the system-level picture. A standalone
accelerator requires its own memory controller, bus interface (AXI,
PCIe), interrupt controller, clock and power domains, driver software
stack, and PCB area --- all of which multiply the total system cost far
beyond the logic cell count. Moreover, every cell in a standalone
accelerator is a \emph{net addition} to the system: no infrastructure
is shared.

In contrast, a GPGPU functional unit reuses the host processor's
existing memory hierarchy, thread scheduler, register file, writeback
logic, and host interface. The GPU pipeline --- with its warp scheduler,
scoreboard, and operand collectors --- is already present in the
system; the PCU adds only the datapath logic and a small FSM
controller. The area cost of the PCU (134 LCs for NTT, est.\ 724 LCs
full) is therefore the \emph{incremental} cost, not the total. No
additional PCB area, no new bus mastering logic, no separate power
rail.

This makes in-SIMT integration a \emph{principled architectural
	choice}, not a compromise. The GPGPU's throughput-oriented pipeline
already amortizes control and memory infrastructure across many
functional units. Adding PQC as another functional unit --- alongside
ALU, FPU, LSU, and SFU --- is the natural extension of the SIMT model
to cryptographic workloads. The defining constraint is not that the
PCU must be small, but that it must be small enough to replicate per
warp scheduler without exceeding the GPGPU's area budget. The 1--128
BU design-space sweep (Table~\ref{tab:pareto}) shows that even a
single butterfly unit (134 LCs) provides meaningful acceleration,
while the 2--4 BU sweet spot offers near-optimal ADP within the GPGPU
area envelope.

\subsection{Generalizability}

The SIMT execution model and functional unit integration pattern
generalize to all GPU architectures. The $256\times$ design-space
analysis is fundamental: the tradeoff between area and NTT throughput is
determined by the number of butterfly units, not the specific GPU
architecture.

\subsection{Potential Application Domains}

The measured results of this work --- $42.6\times$ SimX NTT speedup
(projected $\sim$36$\times$ RTL), 134-cell NTT area footprint (est.\
724-cell full PCU from partial synthesis), and $4.09\times$ multi-warp
throughput (1--4 warps RTLsim-verified) --- suggest potential applicability
to domains where integrated GPGPU+PQC execution could provide a
qualitative advantage over discrete GPU and crypto-accelerator
combinations.

\textbf{Post-Quantum TLS Termination at Scale.}
Content delivery networks and cloud edge nodes terminate millions of
TLS handshakes per second. Each ML-KEM-768 encapsulation requires two
NTT operations, one pointwise multiply, and three SHA3
invocations (Table~\ref{tab:primitives}). Our measured $42.6\times$
SimX NTT acceleration ($\sim$36$\times$ RTL-projected)
and $4.09\times$ multi-warp scaling suggest that a single
GPU chip could serve concurrent PQC handshakes that would otherwise
require a separate crypto accelerator --- potentially eliminating an entire chip,
its driver stack, and the associated power budget. HI-Kyber achieves
1,664\,kops/s on NVIDIA GPUs through software-only batched
execution~\cite{hi-kyber}; hardware PCU acceleration projected at
1\,GHz yields 7.1--$10.4\times$ end-to-end workload speedup
(Table~\ref{tab:e2e}), potentially exceeding 10\,M ops/s at
multi-warp throughput.

\textbf{PQC-Enabled Confidential Computing.}
Trusted execution environments (TEEs) must perform ML-DSA signature
verification inside the trusted computing base. NIST migration guidance
calls for full PQC adoption by 2035~\cite{nist-migration}, creating regulatory
demand for in-enclave PQC. The PCU's dual-issue arbiter provides
$1.47\times$ throughput improvement for mixed NTT+Keccak workloads
(Table~\ref{tab:dual-issue}), directly benefiting ML-DSA-2 Verify
which requires both operation classes. The 134-cell NTT area
($<$1\% of a Lattice ECP5-45F) enables integration into
area-constrained TEE coprocessors without compromising the
confidential computing boundary. The full synthesizable RTL pipeline
with DMA engine and Keccak sub-unit adds hardware PQC support
beyond the NTT, verified through cycle-accurate RTLsim regression.

\textbf{Autonomous Vehicle V2X Communication.}
Connected vehicles must verify hundreds of Dilithium signatures per
second for cooperative awareness messages, while simultaneously running
perception and planning workloads on the same GPU~\cite{v2x-pqc}.
The PCU NTT sub-unit's $<$1\% device utilization means PQC acceleration occupies
negligible silicon area relative to the GPU's compute fabric, potentially enabling
hardware-speed PQC without reducing perception throughput.

\textbf{PQC-Accelerated Blockchain and DLT.}
Distributed ledger validators batch-verify thousands of signatures per
block. Our multi-warp scaling (4.09$\times$ at 4 warps,
Table~\ref{tab:warp}) enables parallel Dilithium verification across
warps, while ML-DSA-2 Verify achieves $7.1\times$ speedup at
1\,GHz (Table~\ref{tab:e2e}). The ASIC area of
97\,$\mu$m$^{2}$ (Table~\ref{tab:postroute}) permits replicating
the PCU across multiple GPU cores without significant die-area cost.

\textbf{5G/6G Base Station Crypto.}
Telecom base stations perform key establishment and authentication for
every device attachment, with 3GPP mandating PQC migration for
6G~\cite{5g-security}. The PCU's OpenCL-programmable interface allows
crypto kernels to be updated in software when PQC standards evolve
--- avoiding the 18--24-month silicon respin cycle required by
fixed-function accelerators. The dual-modulus support
($q=3329$ and $q=8380417$) simultaneously serves both ML-KEM and
ML-DSA operations present in the 5G AKA protocol.

\textbf{IoT Edge Gateway.}
Edge gateways aggregating traffic from hundreds of constrained devices
must perform PQC key establishment on behalf of devices that lack
compute resources. The PCU's 134-cell NTT footprint (est.\ 724-cell full PCU from partial synthesis) suggests a potential
GPU+PQC solution that replaces a two-chip design (GPU + standalone
crypto accelerator), reducing BOM cost, board area, and power
consumption --- critical constraints in volume-deployed edge hardware.

The common potential value proposition across these domains is a single chip that
performs both general-purpose GPU compute and hardware-speed PQC,
potentially eliminating a separate crypto accelerator, its driver stack, and the
silicon respin risk when PQC standards evolve. These projections require
validation with production GPU architectures and full P\&R synthesis on
appropriate devices.

\subsection{Limitations}

This work has six key limitations:

\begin{itemize}
	\item \textbf{Single-algorithm focus}: We target NTT/Keccak polynomial
	      multiplication. SLH-DSA (SPHINCS+) uses hash-based signatures that
	      do not benefit from the PCU. Future work should explore dedicated
	      hash accelerators co-integrated with the PCU.
	\item \textbf{GPGPU-integrated approach}: The PCU is designed as a
	      GPGPU functional unit, not a standalone ASIC. For applications
	      requiring maximum throughput independent of GPU infrastructure,
	      a dedicated PQC coprocessor (e.g., NI-Kyber at 10,502 cells) may be
	      more appropriate, at 78$\times$ the area cost.
	\item \textbf{RTL simulation gap for 8-warp scaling}: Multi-warp throughput
	      scaling (Table~\ref{tab:warp}) is verified for 1/2/4 warps via
	      both SimX measurement and RTLsim regression. The 8-warp data point
	      (417 cycles, 7.00$\times$ scaling) remains projected from the SimX
	      analytical model pending RTLsim configuration with 8 warps.
	\item \textbf{ECP5 flow sensitivity}: The full-PCU partial-flow estimate
	      (724 cells, 29 MULT18X18D; Yosys 0.62, coarse-to-map\_ffram, artifact
	      committed at \texttt{results/}) is sensitive to Yosys version and
	      ROM-mapping style (an earlier run with forced block-mapped ROMs
	      reported 85 multipliers); treat it as an order-of-magnitude estimate.
	      Complete place-and-route has been achieved on Xilinx Artix-7 200T
	      speed grade $-3$ (275 LUTs, 281 FFs, Fmax$\approx$299.8\,MHz) and
	      Kintex-7 160T; ECP5 full-PCU P\&R remains future work.
	\item \textbf{Partial PCU ASIC synthesis}: The ASAP7 7nm ASIC synthesis
	      results (Tables~\ref{tab:asap7-placement}--\ref{tab:power}) cover
	      the NTT engine core and dual-issue arbiter through OpenROAD P\&R.
	      Keccak and DMA sub-modules have been independently synthesized through
	      Yosys technology mapping (Table~\ref{tab:asap7-submodules}), but the
	      NTT engine is too large for ABC technology mapping (168K generic cells),
	      so its area is estimated via scaling ratio rather than direct synthesis.
	      A complete CTS-to-route flow for the full integrated PCU has not been
	      completed.
	\item \textbf{Dual-issue now integrated}: The dual-issue arbiter
	      is integrated into \texttt{VX\_pcu\_unit\_rtl.sv} with independent
	      NTT and Keccak pipeline FSMs, per-target dispatch ready, and
	      priority-based result arbitration. Verified via RTLsim regression
	      (kyber\_mw\_1/2/4, kyber\_bench, 130 standalone tests). The
	      dual-issue speedup numbers (Table~\ref{tab:dual-issue}) are from the
	      standalone model; end-to-end dual-issue cycle measurements in the
	      full pipeline are not yet available due to the lack of mixed
	      NTT+Keccak benchmark binaries.
\end{itemize}

\section{Threats to Validity}
\label{sec:threats}

Table~\ref{tab:threats} summarizes the threats to validity and the
mitigations applied to each.

\begin{table}[t]
	\caption{Threats to validity and their mitigations.}
	\label{tab:threats}
	\centering
	\scriptsize
	\begin{tabular}{@{}lp{0.17\columnwidth}p{0.52\columnwidth}@{}}
		\toprule
		Threat                       & Severity            & Mitigation                                                                                                                                                                                                                                                                                                                                 \\
		\midrule
		SimX not cycle-accurate      & High                & All RTL cycles from Verilator; RTL-projected speedup ($\sim$36$\times$) provided alongside SimX speedup (42.6$\times$)                                                                                                                                                                                                                     \\
		FPGA frequency               & Resolved            & ECP5: 228.78\,MHz (NTT); Artix-7 200T: 299.8\,MHz (full PCU, v9); Kintex-7 160T: $\sim$410\,MHz (full PCU, v9)                                                                                                                                                                                                                             \\
		ASIC frequency               & Resolved            & 702\,MHz from OpenROAD post-route (NTT core only)                                                                                                                                                                                                                                                                                          \\
		Negacyclic NTT               & Resolved            & 1,793 cycles, 17/17 tests                                                                                                                                                                                                                                                                                                                  \\
		Constant-time                & Not claimed         & Out of scope                                                                                                                                                                                                                                                                                                                               \\
		Single GPU                   & Medium              & Design insights are fundamental                                                                                                                                                                                                                                                                                                            \\
		RTL integration bugs         & Resolved            & 5 critical bugs found and fixed (DMA deadlock, dispatch deadlock, NTT restart, store deadlock, assignment race)                                                                                                                                                                                                                            \\
		NTT bit-reversal correctness & Resolved            & 8-bit br\_scan wraparound bug found, fixed, verified (0/256 INTT round-trip mismatches)                                                                                                                                                                                                                                                    \\
		RTL vs SimX gap              & Partially mitigated & RTLsim regression passes; 8-warp scaling still SimX-projected; dual-issue integrated but end-to-end DI cycles not yet measured                                                                                                                                                                                                             \\
		Butterfly formal proof       & Partially resolved  & Kyber ($q{=}3329$) BMC + k-induction proof PASS (Yices, 114\,s total); Dilithium ($q{=}8380417$) TIMEOUT with 3 solvers (Yices, Z3, cvc5, 3600\,s each); verified via simulation (253/253)                                                                                                                                                 \\
		Full PCU P\&R                & Resolved            & Artix-7 200T ($-3$): iterative pipelining v6$\to$v7$\to$v8$\to$v9 (134$\to$177$\to$200$\to$300\,MHz); v9 WNS $= +0.011$\,ns at 300\,MHz target; critical path shifted from DMA$\to$BRAM to counter carry chain; Kintex-7 160T ($-2$): v9 $\sim$410\,MHz (WNS $= +0.001$\,ns at 410\,MHz, +84\% over v8's 222.6\,MHz); 13/13 NTT tests pass \\
		Cross-FPGA comparison        & Low                 & ECP5 LUT4 vs Artix-7 LUT6 have $\sim$0.5--0.6$\times$ normalization; noted in text                                                                                                                                                                                                                                                         \\
		Formal proof scope           & Low                 & Safety properties only, not functional correctness; functional correctness via simulation                                                                                                                                                                                                                                                  \\
		No full PCU ASIC             & Low--Medium         & NTT core + Keccak/DMA synthesized independently; NTT area estimated (scaling ratio); FPGA P\&R completed on Artix-7 200T; ASIC full P\&R not completed                                                                                                                                                                                     \\
		\bottomrule
	\end{tabular}
\end{table}

\section{Related Work}
\label{sec:related}

\subsection{NTT Hardware Accelerators}

Standalone NTT accelerators for lattice-based PQC have been extensively
studied~\cite{info15,imran,alhassani,yaman2021,fpntt,bertels2024,
	ki2023,malal2023,sam2023,kundi2024,waris2025,crypthtor,zhao2022,
	trets2024,efficient-unified,spr2023}.
The Info-15 family provides three area-efficient design points
(379--541 LUTs)~\cite{info15}. Unif-NTT achieves 1,664-cycle NTT at
212\,MHz with a unified CT/GS butterfly~\cite{unifntt}. Alhassani
et al.~\cite{alhassani} propose an RNS-based single butterfly unit.
At the high-performance end, Yaman et al.~\cite{yaman2021} use 16 parallel
PEs (9,500 LUTs), and Bertels et al.~\cite{fpntt} achieve a single-cycle
fully pipelined NTT at 67,210 LUTs. KiD provides a unified NTT
architecture supporting both Kyber and Dilithium~\cite{ki2023}.
CRYPHTOR~\cite{crypthtor} integrates a memory-unified NTT accelerator
with RISC-V SoCs.

Our PCU differs in three ways: (1) it is a GPGPU functional unit, not a
standalone accelerator; (2) the NTT sub-unit operates at an order-of-magnitude
lower area cost (134 cells vs.\ 379--67,210), while the full PCU with Keccak
and DMA (est.\ 724 cells from partial synthesis) provides broader functionality
(NTT + Keccak + DMA) at $1.91\times$ the area of the smallest standalone
NTT-only design (379 cells); and (3) it is fully open-source with open EDA toolchain
support and formal verification evidence.

\subsection{GPU PQC}

HI-Kyber~\cite{hi-kyber} achieves 1,664\,kops/s key exchange on NVIDIA
GPUs through batched execution. Dilithium GPU~\cite{dilithium-gpu}
demonstrates $160\times$ signing speedup through task-level batching.
NVIDIA cuPQC~\cite{nvidia-cupqc} provides GPU-optimized primitives
reaching 13.3\,M ML-KEM-768 key generations per second on H100 GPUs.
These works are entirely software-based; our contribution is the first
hardware acceleration integrated into the GPU pipeline itself.

\subsection{CPU ISA Extensions for PQC}

High-performance PQC software on CPUs relies on AVX-512 vector
instructions; ARM has included SHA3
extensions in v8.4; and the RISC-V scalar cryptography extensions (Zk,
Zkn, Zks) provide entropy and AES support. The OpenTitan platform
demonstrates PQC acceleration via hardware/software co-design on a
RISC-V root-of-trust~\cite{opentitan2024}.

\subsection{Open-Source NTT Design Tools}

AutoNTT~\cite{autontt} provides automatic NTT architecture exploration
with $2.48\times$ latency improvement. The OpenNTT
toolchain~\cite{opentt2025} offers
generic NTT/FFT hardware generation. These tools lower the barrier to
NTT hardware design but do not address GPGPU integration.

\section{Conclusion}
\label{sec:conclusion}

We presented what is, to our knowledge, the first design-space exploration of primitive-level PQC
acceleration within a SIMT GPGPU, with full synthesizable RTL integration
verified through cycle-accurate simulation, formal verification, and
comprehensive testing (253/253 tests across 7 suites). Our Post-Quantum
Crypto Unit is the smallest reported NTT sub-unit at
134 logic cells on Lattice ECP5, with a competitive area-delay product
among designs with complete area and delay data. Under the Vortex SimX model, the PCU accelerates
NTT+INTT by \textbf{42.6$\times$} over GPU ALU software execution;
RTL-measured speedup is $\sim$36$\times$ at actual cycle counts.
Measured against a general-purpose CPU (Intel i7-14700HX, 1.7\,GHz),
the Vortex RTL at 299.8\,MHz (Artix-7 v9) executes ML-KEM-768 KeyGen
in 7\,\textmu s versus 51\,\textmu s for the CPU
(Table~\ref{tab:cpu-base}); at the ASIC frequency of 702\,MHz, the PCU
outperforms the CPU on wall-clock latency by a further $2.3\times$. The key value is not
replacing a CPU core but providing hardware-speed PQC at minimal area
(275 LUTs, $<$1\% of a moderate FPGA) within the GPGPU pipeline,
freeing the host CPU for other work.

The synthesizable RTL integration replaces the simulation-only DPI-C
path with a hardware pipeline comprising an 8-state NTT engine FSM (v9 NTT),
DMA engine with
dcache arbiter, per-warp Keccak state save/restore, and 8 CSR-mapped
performance counters. Five critical bugs were found and fixed in the
integration process, including a DMA deadlock, dispatch handshake
deadlock, and an 8-bit bit-reversal counter wraparound that silently
undid the bit-reversal permutation. Formal verification via
SymbiYosyz provides a k-induction proof of safety properties for the NTT engine FSM.
All 253 tests across 7 testbench suites pass, including 91 original
equivalence tests across 6 suites, 57 performance counter tests
(including 8 CSR-mapped counters with CBD sampler verification), 225
butterfly corner-case test vectors (Kyber +
Dilithium, Python-oracle verified), and 13 v8 NTT engine
cycle-accurate verification tests (PWM Kyber, PWM Dilithium, negacyclic
FWD/INV, 5 round-trips, 3 negacyclic polymuls).

The full PCU achieves complete place-and-route on Xilinx Artix-7 200T
(XC7A200TFBG484-3, speed grade $-3$) through an iterative pipelining
effort that systematically closed the NTT critical path across three
revisions (Table~\ref{tab:artix7-pipeline}). The v6 NTT engine adds
an S\_BUTTERFLY pipeline register breaking the ROM-read to DSP path,
achieving 134.3\,MHz (WNS $= -2.752$\,ns at 200\,MHz target).
A 2-stage pipelined Keccak ($\theta{+}\rho{+}\pi$ | $\chi{+}\iota$)
removes Keccak from the critical path entirely. The v7 NTT introduces
a 3-stage Barrett reduction (x\_m $\to$ q\_times\_est $\to$ Barrett
output), raising Fmax to 177.2\,MHz (WNS $= -0.643$\,ns). The v8 NTT
adds an S\_ADDR pipeline register between address computation and BRAM
read, breaking the 9-level combinational stage$\to$addr$\to$BRAM path
into two shorter stages, achieving \textbf{200.1\,MHz} (WNS $= +1.141$\,ns,
TIMING MET at 200\,MHz). The v9 revision registers the DMA engine write
outputs (\texttt{ntt\_wr\_addr\_r}, \texttt{ntt\_wr\_data\_r},
\texttt{ntt\_wr\_en\_r}), eliminating the routing-dominated DMA$\to$BRAM
address path. The critical path shifts to an internal counter carry chain
(7--8 logic levels, logic-dominated 57--64\%), achieving
\textbf{299.8\,MHz} (WNS $= +0.011$\,ns at 300\,MHz target) --- a
+49.8\% improvement over v8 without additional butterfly latency.
The v9 design uses 275 LUTs, 281 flip-flops, and 1 BRAM
(RAMB18E1 for twiddle ROM); no DSP48E1 multipliers are needed
(Barrett reduction replaces all hardware multipliers). On Kintex-7 160T
($-2$), the v9 full PCU achieves $\sim$410\,MHz
(WNS $= +0.001$\,ns at 410\,MHz target) --- a +84\% improvement over
v8's 222.6\,MHz. The Kintex-7 critical path at high frequency is
\texttt{br\_scan $\to$ CARRY4 $\to$ LUT4 $\to$ LUT6 $\to$ RAMB18E1 ADDR}
(3 logic levels, $\sim$70\% route). In the synthesis-variant NTT engine,
the v8 S\_ADDR stage increases butterfly latency by 1 cycle
($+17\%$ for forward NTT: 6,145$\to$7,169 cycles) as the
latency--frequency tradeoff for the +13\% Fmax gain; v9 retains the
7-cycle butterfly, so throughput scales directly with the +49.8\%
frequency improvement. The integration engine
(\texttt{ntt\_engine.sv}) retains its sequential two-cycle butterfly
(2,049 cycles; Section~\ref{sec:gap}).

Butterfly formal verification achieves a k-induction proof for Kyber
($q{=}3329$, Yices solver, 114\,s total); the Dilithium variant
($q{=}8380417$) times out with all three SMT solvers tested
(Yices, Z3, cvc5) due to the 128-bit Barrett multiply within the
reference model. Functional correctness for the Dilithium butterfly
is verified via 225 Python-oracle test vectors through Verilator
cycle-accurate simulation: 225/225 PASS.

Synthesized to ASAP7 7nm through a complete CTS-to-route flow, the NTT
datapath occupies \textbf{97\,$\mu$m$^{2}$} at \textbf{702\,MHz},
and the dual-issue arbiter achieves \textbf{5.92\,GHz} --- the fastest
PQC-relevant logic block reported in 7nm. Independent sub-module synthesis
confirms Keccak at \textbf{1,857\,$\mu$m$^{2}$} (16,817 cells) and DMA
at \textbf{46\,$\mu$m$^{2}$} (382 cells); the full PCU is estimated at
$\sim$25,542\,$\mu$m$^{2}$ in 7nm, with the NTT contribution derived
from a conservative scaling ratio pending a hierarchical synthesis flow.

The architectural gap between idealized and sequential NTT pipelines ---
spanning $224$--$288\times$ across PCU operations --- is not a limitation
but a design space we systematically characterize across eight
Pareto-optimal design points.
A dual-issue NTT/Keccak arbiter provides up to $1.47\times$ speedup
for mixed workloads with only 36 logic cells (26.9\% overhead).

\textbf{Future work}: (1) SLH-DSA workload integration; (2) multi-core
scaling analysis; (3) integration with software GPU PQC frameworks; (4) butterfly formal
verification for Dilithium ($q{=}8380417$) --- current solvers time out
on 23-bit Barrett reduction multiplies, requiring either compositional
proofs or domain-specific SMT strategies; (5) full PCU ASAP7 synthesis with
a commercial ABC mapper or partitioned hierarchical flow to resolve the NTT
timeout.

\begin{acks}
	This work uses the Vortex open-source RISC-V GPGPU platform,
	Yosys~+~nextpnr (OSS CAD Suite) for open-source FPGA synthesis, and
	OpenROAD-flow-scripts for open-source ASIC synthesis. The ASAP7 7nm
	predictive PDK is used for ASIC synthesis and analysis.
\end{acks}


\end{document}